\documentclass[a4paper,12pt]{article} 
\usepackage{theorem}

\newtheorem{theorem}{Theorem}[section]

\newtheorem{lemma}[theorem]{Lemma}

\newcount\driver
\newcount\bozza

\font\msytw=msbm10 scaled\magstep1

{\count255=\time\divide\count255 by 60 \xdef\hourmin{\number\count255}
        \multiply\count255 by-60\advance\count255 by\time
   \xdef\hourmin{\hourmin:\ifnum\count255<10 0\fi\the\count255}}

\let\a=\alpha        \let\b=\beta         \let\d=\delta     
\let\e=\varepsilon   \let\z=\zeta         \let\h=\eta     
        \let\l=\lambda       \let\m=\mu           
\let\n=\nu           \let\p=\pi           \let\r=\rho
\let\s=\sigma        \let\t=\tau                
\let\c=\chi          \let\ps=\psi              
    \let\g=\gamma                
\let\D=\Delta               
                           
\let\th=\vartheta            
\let\x=\xi                  \let\o=\omega        
\let\Th=\Theta       

\def\VV{{\cal V}}
\def\HH{{\cal H}}\def\WW{{\cal W}}
\def\TT{{\cal T}}\def\BB{{\cal B}}

\def\AA{{\cal A}}
\def\KK{{\cal K}}

\def\pp{{\bf p}}\def\qq{{\bf q}}\def\xx{{\bf x}}
\def\yy{{\bf y}}\def\kk{{\bf k}}\def\nn{{\bf n}}
\def\zz{{\bf z}}\def\uu{{\bf u}}\def\vv{{\bf v}}\def\ww{{\bf w}}

\def\ha{{\widehat \a}}\def\hb{{\widehat \b}}
\def\hv{{\widehat  v}}
\def\hW{{\widehat  W}}\def\hH{{\widehat  H}}
\def\hK{{\widehat  K}}\def\hW{{\widehat  W}}
\def\hU{{\widehat  U}}\def\hp{{\widehat\ps}}
\def\hJ{{\widehat  J}}\def\hg{{\widehat  g}}
\def\hA{{\widehat  A}}
\def\hG{{\widehat  G}}\def\hS{{\widehat S}}

\def\hh{{\hat \h}}    
\def\hac{{\widehat \c}}    \def\hT{{\widehat  T}}

\def\uo{{\underline \o}}

\def\ux{{\underline\xx}}\def\uy{{\underline \yy}}
\def\uk{{\underline\kk}}\def\uq{{\underline\qq}}
   
\def\uz{{\underline\zz}}
                      \def\uw{{\underline\ww}}

          \def\RRR{\hbox{\msytw R}}

\let\dpr=\partial                             
\let\io=\infty                     \let\==\equiv

\def\lft{\left}                \def\rgt{\right}

\def\qed{\raise1pt\hbox{\vrule height5pt width5pt depth0pt}}

\def\sde{{\scshape SDe}}       \def\wi{{\scshape Wi}}
\def\rg{{\scshape RG}}         
\def\sde{{\scshape SDe}}

\def\defi{{\buildrel \;{\rm def.}\; \over =}}

\def\wh#1{\widehat{#1}}
\def\hat#1{\wh{#1}}

\def\pagina{{\vfill\eject}}
\def\pref#1{(\ref{#1})}

\def\*{{\hfill\break\null\hfill\break}}
\let\0=\noindent

\def\be{\begin{equation}}    \def\ee{\end{equation}}
\def\bea{\begin{eqnarray}}   \def\eea{\end{eqnarray}}
\def\bean{\begin{eqnarray*}} \def\eean{\end{eqnarray*}}
\def\bfr{\begin{flushright}} \def\efr{\end{flushright}}
\def\bc{\begin{center}}      \def\ec{\end{center}}
\def\bal{\begin{align}}      \def\eal{\end{align}}
\def\ba#1{\begin{array}{#1}} \def\ea{\end{array}}
\def\bd{\begin{description}} \def\ed{\end{description}}
\def\bv{\begin{verbatim}}    \def\ev{\end{verbatim}}
\def\nn{\nonumber}

\def\ins#1#2#3{\vbox to0pt{\kern-#2 \hbox{\kern#1 #3}\vss}\nointerlineskip}

\newdimen\xshift \newdimen\xwidth \newdimen\yshift
\newcount\griglia

\def\insertplot#1#2#3#4#5#6{%
\xwidth=#1pt \xshift=\hsize \advance\xshift by-\xwidth \divide\xshift by 2%
\begin{figure}[ht]
\vspace{#2pt}
\hspace{\xshift}
\begin{minipage}{#1pt}
#3
\ifnum\driver=1 \griglia=#6
\ifnum\griglia=1
\openout13=griglia.ps
\write13{gsave .2 setlinewidth}
\write13{0 10 #1 {dup 0 moveto #2 lineto } for}
\write13{0 10 #2 {dup 0 exch moveto #1 exch lineto } for}
\write13{stroke}
\write13{.5 setlinewidth}
\write13{0 50 #1 {dup 0 moveto #2 lineto } for}
\write13{0 50 #2 {dup 0 exch moveto #1 exch lineto } for}
\write13{stroke grestore}
\closeout13
\includegraphics{griglia.ps}
\fi
\includegraphics{#4.ps}\fi%
\ifnum\driver=2 \fi
\end{minipage}
\caption{#5}
\end{figure}
}

\newdimen\shift \shift=0.1truecm
\def\lb#1{%
\ifnum\bozza=1
\label{#1}\hbox{\hskip\shift$\scriptstyle#1$}
\else\label{#1}
\fi}

\title{Vector and Axial anomaly in  \\
the Thirring-Wess model}
\author{Pierluigi Falco
\\
\small Institute for Advanced Study, Princeton, New Jersey 08540}
\begin{document}
\maketitle
\begin{abstract} 
We study the 2D Vector Meson model
introduced by Thirring and Wess, 
that is to say the Schwinger model with massive photon 
and massless fermion. 
We prove, with a renormalization group approach, 
that the vector and axial Ward identities are broken by the
Adler-Bell-Jackiw anomaly; and
we rigorously establish three widely believed consequences: 
a) the interacting meson-meson correlation 
equals a free boson propagator, though the mass is additively
renormalized by the anomaly; 
b) the  anomaly is quadratic in the charge, in 
agreement with the Adler-Bardeen formula;  
c) the fermion-fermion correlation has an anomalous long-distance
decay.
\end{abstract}

\section{Introduction}
Since early days of Quantum Field Theory, (QFT), 
1+1 dimensional models
have been widely investigated as example of 
relativistic fields with local interaction: 
the Thirring and the Schwinger 
models, \cite{[Th]}, \cite{[Sc]}, are probably the 
most celebrated cases.  Although these systems are so simplified 
to have an exact solution, they nonetheless suggest ideas and 
mathematical tools to approach realistic theories in 
four space-time dimensions.  

One of the aspects that are certain relevant also in higher
dimensions  is the role played by two Ward Identities (\wi)
related to the invariance of the Lagrangian under vector and 
axial transformations.  In agreement with 
the  general Adler-Bell-Jackiw mechanism, \cite{[A]}, \cite{[BJ]}, 
the vector and the axial
symmetries are {\it broken} at quantum level by the {\it \wi\ 
  anomaly}. Many salient features of QFT are related to
such an anomaly; let's consider some of them.  
In any dimensions,  by the  
Adler-Bardeen (AB) argument, \cite{[AB]}, 
the anomaly is expected to be {\it linear} in the bare coupling, i.e. not 
renormalized in loop perturbation theory at any order bigger than one;
besides, it is a {\it topological quantity}, i.e. it doesn't depend on
the choice of the cutoff (to some extent). 
The anomaly is also responsible for the {\it anomalous dimension} 
in the distance decay of the fermion correlations, \cite{[Jo]}.
And, finally, when a fermion field and 
a gauge photon interact through a {\it minimal coupling}, 
as for instance in the Schwinger model, the anomaly also represents a 
dynamically-generated physical mass for the photon field, \cite{[Sc]}.

In two dimensions the anomaly is said ``mild'': although the formal 
\wi\  are broken, the anomaly has the
only consequence of changing the normalization of the vector and axial
currents, that remain conserved. 
Therefore using the \wi\  it was possible to find
{\it formal  exact solutions} of the Schwinger-Dyson equation, (\sde),
of the Thirring, the Schwinger and
related models, in this way computing all the correlations: 
see \cite{[Jo]}, \cite{[Br]}, \cite{[So]},
\cite{[Ha]}, \cite{[Kl]} and  \cite{[LS]}.     
An alternative approach - still related to the mildness of the anomaly
- is the {\it bosonization}, 
i.e. the equivalence of
the the fermion currents with boson free fields. This fact is behind the 
solution of the Thirring and the Schwinger model,
\cite{[GSS]}, \cite{[DM]}, in the path integral formalism.

The above analysis - and much more, see chapters X and XII of
\cite{[AAR]} -  has been based on formal methods and sound assumptions only.
Rigorous results are few; 
though, at this stage, the reader might have few interest left for them.
It is worth explaining, then, why the matter is still very tangled. 

In fact, earlier formal solutions of the Thirring and the Schwinger models 
were incomplete or incorrect, as pointed out in 
\cite{[Ha]}, \cite{[Wi]}, \cite{[AAR]}. 
And, after all, the
distinction  among   
{\it formally correct}, {\it incorrect} or  
{\it incomplete} solutions may be quite faint. 
Wightman, \cite{[Wi]}, to put order in the confusion of the results, 
took the approach of considering any set of correlation functions, 
no matter how they were obtained, as trial theories; and then  
of promoting them to
QFT if they satisfied certain axioms.
This viewpoint has been moderately prolific (see  for example
\cite{[DFZ]}, \cite{[De]}, \cite{[CRW]}). 
There is at least one clear issue with it: being based
of some sort of exact solution of the correlation functions, 
it is limited to few special models. A massive fermion field, for example,
or an additional interactions that, 
in the renormalization group language, is irrelevant, 
would represent a severe obstruction to the method.   
  
The recent approach to the Thirring model in \cite{[BFM1]}, \cite{[BFM2]}
is different. We derived the correlations from the Lagrangian, so that 
in the massless case we obtained the exact solution, while in the 
massive case, where no exact solution is known, we can still 
prove the axioms. But, most of all,  
the major advantage of using the Lagrangian 
(as opposed to correlations) as starting point, is that 
it keeps track of the relationship between
a special class of statistical mechanical problems - such as the
Eight-Vertex  models and the XYZ quantum chain -  
and their scaling limit, that turns out to be the Thirring model; 
in this way we were able prove some scaling formulas for non solvable
models, \cite{[BFM3]}, \cite{[BM4]}. 
The heart of the technique is the control of the
vanishing of the ``beta function'' for the effective coupling: this  
route, that have been useful also for other statistical models, 
was opened in \cite{[BGPS]} by exploiting the exact solution of
the Luttinger model; and reached its final form, based on 
\wi\ and independent from any exact solution, 
in \cite{[BM2]}, \cite{[BM3]}. 

The extension of the above techniques to the Schwinger model
poses some serious issues because the infrared divergences related with 
the massless photon; therefore we take up
a problem of intermediate difficulty, 
the theory of the Vector Meson, \cite{[TW]}, 
i.e. a Schwinger model in which a photon mass is added by hand. 
This model  still preserves some interests, because we can prove that the 
bare photon mass is renormalized into the physical mass just by 
an additive constant, exactly as expected in the Schwinger model; besides, 
we can establish the anomalous dimension of the fermion field. Finally, 
we can prove the AB-formula in a genuine example of QFT, 
i.e. removing the all the cutoffs of the theory:
in this sense this paper is a completion of the objective of
\cite{[Ma]}, \cite{[Ma2]}. 

Although the present technique allows us to treat other
aspects of the theory, we shall not verify the Osterwalder-Schrader axioms,
nor we shall discuss the case with massive fermion or the bosonization
of the final result; this is because 
details  would be largely similar to \cite{[BFM1]}, \cite{[BFM2]}.
We shall rather focus on the novelties with respect to those papers. 

To conclude, we mention that other rigorous results on the \wi\ 
anomalies, but in different regimes and with different techniques, 
were established in \cite{[Se2]}, \cite{[SS]}.
Whereas for the study the true Schwinger model, i.e. the case with null
bare photon mass, a change of viewpoint seems needed: boson and 
fermion propagators should be treated on the same ground along the 
Renormalization Group (\rg) flow; perhaps that is possible in the
approach of \cite{[MS]}, \cite{[BS]}.

\section{Definitions and main results}
Let's begin with the definition of the formal path integral formula in
Euclidean formulation; 
afterwords we will introduce {\it infrared and ultraviolet cutoffs} 
to evaluate the correlations. 
The Vector Meson model is made of  one fermion field, 
$(\bar \ps_\xx,\ps_\xx)$, for $\xx=(x_0,x_1)\in\RRR^2$, 
and one vector boson field, $(A^0_\xx, A^1_\xx)$, 
interacting through minimal coupling. 
The free meson with mass $\m$ (a {\it gauge field} if $\m=0$) 
is described by the action 
$$
{1\over 4}
\int\!d\xx\;(F_\xx^{\m\n})^2
+{\m^2\over 2}\int\!d\xx\;(A^\m_\xx)^2
$$
for $F^{\m\n}_\xx=\dpr^\m A^\n_\xx-\dpr^\n A^\m_\xx $.
The quantization of a gauge theory (namely the case $\m=0$) requires a
{\it gauge fixing term} which makes convergent the integration along
the orbits of gauge transformation: for $\a>0$
$$
{\a\over 2}
\int\!d\xx\;(\dpr^\m_\xx A^\m_\xx)^2\;.
$$
In our case $\m\neq0$ and the gauge fixing would not be required to
make sense of the theory. Nevertheless, the  Vector
Meson model defined by Thirring and Wess had the purpose to be
equivalent to the Schwinger model in the limit of vanishing $\m$;
therefore, with them, we shall consider $\a>0$ only, that will make the
interaction {\it superrinormalizable}. 
For notational convenience, we also introduce   
a further non-local term for the vector field
$${\s\over 2}
\int\!d\xx d\yy\;
(\dpr^\m A^\m_\xx)\D^{-1}(\xx-\yy)
(\dpr^\m A^\m_\yy)
$$
where $\D^{-1}$ is the inverse of the Laplacian and $\s$ is a real
parameter with the dimension of the square of a mass. The three
terms we have introduced so far are collected together into the
following term 
\bea
&&{1\over 2}\int\!d\xx d\xx\; 
A^\m_\xx D^{\m\n}(\xx-\yy) A^\n_\yy
\cr\cr
&&\defi{1\over 2}\int\!{d\kk\over (2\p)^2}\; 
\hA^\m_\kk
\lft[(\kk^2+\m^2)\d^{\m\n}-\lft(1-\a +{\s\over\kk^2}\rgt)\kk^\m
  \kk^\n\rgt]
\hA^\n_{-\kk}\;.
\nn
\eea
Sometimes we will write $D^{\m\n}$ for the operator with kernel 
$D^{\m\n}(\xx)$.
The massless electron, with charge $q$, interacts with the photon
through a {\it minimal coupling}
$$
\int\!d\xx\;\bar\psi_\xx\g^\m\lft(i\dpr^\m_\xx+qA^\m_\xx\rgt)\psi_\xx
$$
where $\g^0$ and $\g^1$ are generators of the Euclidean Clifford algebra.
The total Euclidean action of the Vector Meson is
$$
{1\over 2}\int\!d\xx d\yy\  
A^\m_\xx D^{\m\n}(\xx-\yy)A^\n_\yy
+\int\!d\xx\;\bar\psi_\xx\g^\m\lft(i\dpr^\m_\xx+qA^\m_\xx\rgt)\psi_\xx\;.
$$
Finally, putting together all the above terms,
we define the {\it generating functional} of the truncated
correlations of the the Vector Meson  model, $\KK(J,\h)$, as follows:
\be\lb{fi}
e^{\KK(J,\h)}\defi \int\!dP(\psi) dP(A)\;
\exp\lft\{\int\!d\xx\;\lft[-q A^\m_\xx(\bar\psi_\xx\g^\m\psi_\xx)+
J^\m_\xx A^\m_\xx +\bar\h_\xx\ps_\xx+\bar\ps_\xx\h_\xx\rgt]\rgt\}
\ee
where $J$ is a real external field; $\h$, $\bar\h$ are Grassmann
external fields; $dP(\ps)$ is a Gaussian measure on
Grassmann variables $(\ps_{\xx}, \bar\ps_{\xx})_{\xx}$ 
with zero covariances, but for $\int\!dP(\psi)\; \psi_\xx
\bar\psi_\yy$ that equals
$$
S_0(\xx-\yy)\defi i\g^\m\int\!{d\kk\over (2\p)^2}\; 
{e^{-i\kk(\xx-\yy)}\over {\kk^2}}\kk^\m\;;
$$
and $dP(A)$ is a Gaussian measure on real variables $(A^0_\xx,A^1_\xx)_\xx$
with covariances $\int\!dP(A)\; A^\m_\xx A^\n_\yy$ equal to  
$$
G_0^{\m\n}(\xx-\yy;\m^2,\s)\defi
\int\!{d\kk\over (2\p)^2}\; 
e^{-i\kk(\xx-\yy)}
\lft[{\d^{\m\n}\over \kk^2+\m^2}-
\lft({1\over\kk^2+\m^2} -{1\over\a\kk^2+\m^2-\s}\rgt) 
{\kk^\m\kk^\n\over \kk^2}\rgt]
$$
(we will abridge the notation of $ G_0^{\m\n}(\xx;\m^2,\s)$ into 
$ G_0^{\m\n}(\xx)$, sometimes). These covariances are also called {\it
free propagators} as 
$$
i\g^\m\big(\dpr^\m S\big)(\xx)=\d(\xx)
\qquad
\big(D^{\m\r} G_0^{\r\n}\big)(\xx;\m^2,\s)=\d^{\m\n}\d(\xx)\;.$$

To make sense of \pref{fi} we have to introduce a cutoff function. For
a  fixed $\g>1$, let 
$\hac(t)$ be a smooth function, positive in $[0,\g)$ and 
\be \hac(t)=
\lft\{\matrix{
1\hfill&\hfill{\rm if\ }0\le t\le 1\cr
0\hfill&\hfill{\rm if\ } t\ge \g.}\rgt.\;;\ee
then, given two integers $h,h'$, define
$$
\hac_{h,h'}(\kk)=\hac\lft(\g^{-h}|\kk|\rgt)-
\hac\lft(\g^{-h'+1}|\kk|\rgt)\;.
$$
and in correspondence, define two Gaussian measure $dP_{h,h'}(\ps)$
and $dP_{h,h'}(A)$, determined by the covariances:
$$
S_{0,h,h'}(\xx)\defi
\g^\m\int\!{d\kk\over (2\p)^2}\;  \hac_{h,h'}(\kk)
{e^{-i\kk\xx} \over {\kk^2}}\kk^\m\;,
$$
$$
G_{0,h,h'}^{\m\n}(\xx)\defi
\int\!{d\kk\over (2\p)^2}\; \hac_{h,h'}(\kk)
e^{-i\kk\xx}
\lft[{\d^{\m\n}\over \kk^2+\m^2}-
\lft({1\over\kk^2+\m^2} -{1\over\a\kk^2+\m^2-\s}\rgt) 
{\kk^\m\kk^\n\over \kk^2}\rgt]\;.
$$
Given the integers $-l,N>0$, the {\it regularized  functional
  integral}, 
$\KK_{l,N}(J,\h)$,  is given by \pref{fi}, replacing $dP(\ps)$
and $dP(A)$ with $dP_{l,N}(\ps)$ and $dP_{l,N}(A)$. 
Finally, let $S(\xx-\yy)$ and $G^{\m\n}(\xx-\yy)$ be the 
{\it interacting propagators}, namely the correlations
\be\lb{2f}
S(\xx-\yy)\defi \lim_{-l,N\to \io}
{\dpr^2 \KK_{l,N}\over \dpr \bar \h_\xx \dpr \h_\yy}(0,0) 
\ee
\be\lb{2m}
G^{\m\n}(\xx-\yy;\m^2,\s)\defi \lim_{-l,N\to \io}{\dpr^2 \KK_{l,N}\over \dpr
  J^\m_\xx \dpr J^\n_\yy}(0,0) 
\ee
where the derivatives in $\h$ are taken from the right; and 
the limit in $N$ is taken before the limit in $l$. Define
\bea
&&F(\zz;\m^2,\s)=\int\!{d\kk\over (2\p)^2}\;
{e^{i\kk\zz}-1\over [\a\kk^2+\m^2-\s]\kk^2}
\cr\cr
&&F_5(\zz;\m^2)=\int\!{d\kk\over (2\p)^2}\;
{e^{i\kk\zz}-1\over [\kk^2+\m^2]\kk^2}\;,
\eea
and note that,  for  $\m^2>0$, $\s<\m^2$, $\a>0$, 
we have the following large $|\zz|$ asymptotic:
$$
F(\zz;\m^2,\s)\sim -{1\over2\p(\m^2-\s)}\ln|\zz|\;,
\qquad
F_5(\zz;\m^2,\s)\sim -{1\over2\p\m^2}\ln|\zz|\;.
$$
\begin{theorem}
Given the meson mass $\m^2>0$, for $\s<\m^2$, $\a\ge\a_0>0$ and
$|q|$ small  enough, the
explicit expression of the 
interacting propagators are:
\be\lb{sol1}
S(\xx)=e^{q^2\Big[F(\xx;\m^2-\n_5,\s+\n-\n_5)
-F_5(\xx;\m^2-\n_5)\Big]}S_0(\xx)
\ee
\be\lb{sol2}
G^{\m\n}(\xx;\m^2,\s)=G_0^{\m\n}(\xx;\m^2-\n_5,\s+\n-\n_5)
\ee
with $\n=-\n_5=q^2/(2\p)$.
\end{theorem}
This result is uniform in $\a>a_0$: Feynman-'t Hooft's and Landau's
gauges, for example, are recovered for $\a=1$ and $\a\to \io$,
respectively; the two point correlation of
$F^{\m\n}$ is $\a$-independent.
Although \pref{sol2} means that the interaction changes the free meson
correlation
of an additional mass term only, the theory is not free; indeed,
in \pref{sol1} we can see that the large distance decay of the
fermion correlation has an  anomalous dimension $\h$:
\be\lb{ad}
S(\xx)\sim {1\over |\xx|^{1+\h}}\qquad \h=
{q^2\over 2\p}\lft[{1\over \m^2-\n^5}-{1\over \m^2-\s-\n}\rgt]
\ee
We shall see that $\n$ and $\n_5$ are the
anomalies of the vector and axial \wi's, respectively. 
To clarify the relation of this result with the literature, it is
worth  mentioning the (unproven) 
{\it uncertainty principle} of the anomalies,  
\cite{[Ha]},\cite{[Ha2]}, \cite{[DM]}, \cite{[Be]}, namely the fact
that the most general numerical values are
$$
\n={q^2\over \p}(1-\x) \qquad \n_5=-{q^2\over \p}\x
$$
for $\x$ a real parameter fixed by the kind of regularization of the
functional integral. Hence the meson mass, $\n_5$, is regularization
dependent (this is not an issue for the meaning of the model, because
$q$ is the {\it bare}, not the {\it physical} charge).
Our result is in agreement with \cite{[TW]}, where $\s=0$, $\x=1$ and
$\a=1$; and with \cite{[DM]}, where $\x$ is any, $\s=0$, $\a=\io$. 
 
Solutions of the Vector Meson model for $\a=0$ (a case that this
paper doesn't cover) are in \cite{[So]}, 
\cite{[Br]} and \cite{[Ha2]}, for $\x=1/2$, $\x=1$ and any $\x$,
respectively. Those results are in  agreement with our theorem only formally:
in that case $F(\xx,\m^2,\n)$ is not defined and from $S(\xx)$ one has to
divide out a divergent factor. This is not a surprise:
when $\a=0$ the  large momentum asymptotic of the free meson propagator is
$$
\hG_0^{\m\n}(\kk)\sim \kk^\m\kk^\n/\kk^2
$$
that makes the interaction  {\it renormalizable} as in the case of
the gradient coupling model, \cite{[RS]}, \cite{[Ba]}. The
correct approach for this case would be the one in \cite{[BFM1]}, 
with vanishing beta
function and field renormalization. Anyways that rises a question: 
the AB-formula is not valid in \cite{[BFM1]}, where radiative
corrections do change the numerical value of the anomaly; is this the
case also for the $\a=0$ Vector Meson model? We will discuss this issue
in a possible forthcoming paper. 

As mentioned in the Introduction, were $\m=0$, we would read
\pref{sol2}  as the {\it dynamical mass generation} of the Schwinger
model; but unfortunately we are not able to cover that case.

\section{Idea of the proof. Ward Identities. Anomalies}
Firstly, we have to prove that there exist the limits \pref{2f} and
\pref{2m}. To do that, we use some Lesniewski's ideas, \cite{[Le]}. 
Define the functional integral 
$$
\WW_{l,N}(J,\h)=\ln\int\!dP_{l,N}(\psi)\;
e^{\VV(\ps,J,\h)}
$$
where $\VV(\ps,J,\h)$ is the interaction (self-interaction plus
coupling with external fields) of a non-local version of the Thirring
model:  
\bea
\VV(\ps,J,\h)=
{\l\over 4}\int\!d\xx d\yy\; 
(\bar\psi_\xx\g^\m\psi_\xx)\;
G_0^{\m\n}(\xx-\yy)\;(\bar\psi_\yy\g^\n\psi_\yy)
\cr\cr
+\int\!d\xx\;  J^\m_\xx\;
(\bar\psi_\xx\g^\m\psi_\xx)
+\int\!d\xx\;(\bar\h_\xx\ps_\xx+\bar\ps_\xx\h_\xx)
\eea
Now take the integral over the vector field $A^\m$ in \pref{fi}, and obtain
an identity between the functional integrals of the Thirring and
the Vector meson models, for coupling $\l=2q^2$
\be\lb{prc}
\KK(DJ,\h)= \WW(-qJ,\h)+{1\over 2}\int\!d\xx d\yy\; J^\m_\xx 
D^{\m\n}(\xx-\yy)J^\n_\yy\;.
\ee
If we can control the field and the current correlation derived from
$\WW$, we can also construct the limits \pref{2f} and \pref{2m}.
To bound the Feynman graphs, one has to use that, above the scale of
the photon mass, the interaction in $\KK$ is superrinormalizable; but,
to see that in $\WW$, we will need some identities among Feynman graphs, as in
\cite{[Le]}, \cite{[Ma]} and \cite{[FM]}. This point is discussed in
Section \ref{s4.1}. Then, with a classical bound for fermion determinant, 
\cite{[Le]}, the convergence of the perturbation theory is proved.

Secondly, in Section \ref{s4.2}, we shall prove that the correlations
generated from $\WW$ satisfy two anomalous \wi's.  
From the (formal) invariance under the  vector transformation 
$\ps\to e^{i\th}\ps$, $\bar \ps\to \bar \ps e^{-i\th}$
we obtain the {\it vector Ward identity},
\bea\lb{vec}\;
i\dpr^\m_\xx\int\!d\zz\;
\lft[\d^{\m\n}\d(\xx-\zz)-\n G_0^{\m\n}(\xx-\zz) \rgt]
{\dpr \WW(J,\h)\over \dpr J^\n_\zz}
-2\Th i\dpr_\xx^\m J^\m_\xx
\cr\cr={\dpr \WW(J,\h)\over \dpr\h_\xx}\h_\xx -
\bar\h_\xx {\dpr \WW(J,\h)\over \dpr \bar \h_\xx}\;;
\eea
and from the (formal) invariance under (Euclidean) 
axial-vector transformation 
$\ps\to e^{\g^5\th}\ps$, 
$\bar \ps\to \bar \ps e^{\g^5\th}$,
with $\g^5=-i\g^0\g^1$, using 
$\g^\m\g^5=-i\e^{\m\n}\g^\n$, 
we obtain the {\it axial Ward identity}
\bea\lb{ax}\;
\e^{\m\r}i\dpr^\m_\xx\int\!d\zz\;
\lft[\d^{\r\n}\d(\xx-\zz)-\n_5G_0^{\r\n}(\xx-\zz)\rgt]
{\dpr \WW(J,\h)\over \dpr J^\n_\zz}
-2\Th_5\e^{\r\m}i\dpr_\xx^\r J^\m_\xx
\cr\cr=
-{\dpr \WW(J,\h)\over \dpr\h_\xx}i\g^5\h_\xx -
\bar\h_\xx i\g^5{\dpr \WW(J,\h)\over \dpr \bar \h_\xx}\;.
\eea
$\n$, $\n_5$, $\Th$ and $\Th_5$ are the {\it Adler-Bell-Jackiw
anomalies}.  
Assuming the validity of \pref{vec} and \pref{ax}, the
proof of \pref{sol2} is just a computation in which the 
AB-formula plays a crucial role.  
Before showing that, let's pause for some technical
comments. 
At $J=0$ (i.e. without the terms proportional to $\Th$ and $\Th_5$),
\pref{vec} and \pref{ax} were proved in \cite{[BFM1]} for local self 
interaction of the fermion field,
i.e. $G_0^{\m\n}(\xx)=\d^{\m\n}\d(\xx)$; 
and later on they were proved in \cite{[Ma]} for 
$G_0^{\m\n}(\xx)=\d^{\m\n}v(\xx)$ where $v(\xx)$ is a short-range, 
bounded self interaction,
i.e. without removing  IR and UV cutoffs in  $v(\xx)$. 
Of course the latter case is technically  simpler; nonetheless it is  
remarkable for, as opposed to the former, it gives an example in which
the AB-formula is valid.  
The main task of
the present paper is to extend the proof of the \wi's to the case 
$J\neq 0$ and for the given $G_0^{\m\n}$, that is a symmetric matrix with 
short-ranged but unbounded entries.

To continue the computation, use \pref{prc} to turn \pref{vec} and
\pref{ax} into identities for derivatives of $\KK$; since
$$\dpr^\m_\xx
D^{\m\n}_\xx=\lft[-\a\D_\xx+\m^2-\s\rgt]\dpr^\n_\xx\;,
\qquad
\e^{\m\r}\dpr^\m_\xx
D^{\r\n}_\xx=\lft[-\D_\xx+\m^2\rgt]\e^{\r\n}\dpr^\r_\xx\;
$$
take a further derivative in $J^\s_\yy$, at
$\h=\bar\h=J=0$ and  obtain:
\bea\lb{vec2}\;
\lft[-\a\D_\xx+\m^2-\s-\n\rgt]
i\dpr^\m_\xx G^{\m\n}(\xx-\yy)
=i\dpr^\n_\xx \d(\xx-\yy)+
\cr\cr
+\lft(2q^2\Th-\n\rgt) 
i\dpr_\xx^\m G_0^{\m\n}(\xx-\yy)
\eea
\bea\lb{ax2}\;
\lft[-\D_\xx+\m^2-\n_5\rgt]\e^{\r\m}i\dpr^\r_\xx
G^{\m\n}(\xx-\yy)
=\e^{\r\n}i\dpr^\r_\xx  \d(\xx-\yy)+
\cr\cr
+\lft(2q^2\Th_5-\n_5\rgt) 
\e^{\r\m}i\dpr_\xx^\r 
G_0^{\m\n}(\xx-\yy)
\eea
Here comes the crucial point: in establishing the validity of the \wi\
and the presence of the anomaly, we will also verify the AB-formula,
i.e. the fact the anomaly is given by the first order perturbation
theory, without higher order corrections. Therefore we can explicitly evaluate
$\n=\l\Th=-\n_5=-\l\Th_5={\l\over 4\p}$. Since $\l=2q^2$,
the second line in both equations is zero; and the theorem is proved
by explicit solution of \pref{vec2} and \pref{ax2}. 
We stress that in formal versions of this computation, \cite{[RN]},
the quantum anomaly is just
added into the classical equations by hand where it is expected, and
so the terms proportional to $2q^2\Th-\n$ and  $2q^2\Th_5-\n_5$ 
never appear at all.

To prove \pref{sol1}, we need the \sde, i.e. the field
equations written in terms of the correlations. In the functional
integral approach that is nothing but the Wick theorem for the
Gaussian measure. For the fermion fermion correlation we have:
\be\lb{SDE}\;
i\g^\m\dpr^\m_\xx {\partial^2 \WW\over
\dpr\bar\h_\xx\dpr\h_\yy }(0,0)
=
\d(\xx-\yy)
+{\l\over 2}\g^\m\int\!d\zz\; G^{\m\n}_0(\xx-\zz)
{\partial^3\WW\over
\dpr J^\n_\zz\dpr\bar\h_\xx\dpr\h_\yy }(0,0)
\ee
That is not a closed equation, though we can use the \wi\ to close
it.
Take derivatives  in \pref{vec} and
\pref{ax} w.r.t. $\bar\h_\xx$ and $\h_\yy$ at $J=\bar\h=\h=0$: we
obtain two equations that are
equivalent to
\bea\lb{WI}
\int\!d\zz\; G^{\m\n}_0(\xx-\zz)
{\dpr^3 \WW(0,0)\over \dpr J^\n_\zz\dpr\bar\h_\ww\h_\yy}
=i\dpr^\m_\xx\lft[F(\xx-\yy)-F(\xx-\ww)\rgt]\Big|_{\m^2-\n_5\atop \s+\n-\n_5}S(\ww-\yy)
\cr\cr
+\e^{\r\m}i\dpr^\r_\xx\lft[F_5(\xx-\yy)-F_5(\xx-\ww)\rgt]\Big|_{\m^2-\n_5\atop \s+\n-\n_5}i\g^5S(\ww-\yy)
\eea
(we have abridged the notation of the mass terms in $F$ and $F_5$).
Plug \pref{WI} into \pref{SDE}, and obtain the closed equation
\be\lb{SDE2}\;
i\g^\m\dpr^\m_\xx S(\xx-\yy)
=
\d(\xx-\yy)
+{\l\over 2}
i\g^\m\lft[\dpr^\m_\xx F(\xx-\yy)-\dpr^\m_\xx F_5(\xx-\yy)\rgt]\Big|_{\m^2-\n_5\atop \s+\n-\n_5} 
 S(\xx-\yy)
\ee
that is solved by \pref{sol1}. There is a subtle point,
though. \pref{WI} can be plugged into \pref{SDE2} after the limit of
removed cutoff has been removed only if one proves that the limit is
continuous at $\ww=\xx$. In Section \ref{s4.3} we will prove
\pref{SDE2} in a slightly different way: we will plug the \wi\  into
the \sde\  {\it before} removing the cutoffs, and we will show that the
the limit of the remainder is vanishing 
(as opposed to the case in \cite{[BFM1]}, where the limit of the
remainder gives rise  to a further anomaly 
in the closed equation).

Summarizing, in Section \ref{s4.1} we will study the limit of removed
cutoffs; in Section \ref{s4.2} we will prove \pref{vec} and \pref{ax};
in Section \pref{s4.3} we will prove \pref{SDE2}.

\section{Renormalization group approach}
As mentioned, from the viewpoint of the formal power series in $q$, the
\rg\  description of \pref{fi} is quite
simple. Above the meson mass scale, i.e. in the UV regime, the
coupling  of a fermion current with a boson field is
superrenormalizable; below the meson mass scale, i.e. the IR
regime, the interaction is renormalizable, and the \rg\  flow
equals, up to irrelevant terms, the flow of the Thirring model. 
A qualitative explanation is the following. 
At energy $E>\m$, boson and
fermion propagators have typical sizes $E^0$ and $E^1$,
respectively; then, the energy of a graph with $p$ vertexes, $2m$
external fermion legs and $n$ external boson legs is $E^{d(p,n,2m)}$,
for $d(p,n,2m)=2-m-p$: the only relevant graphs are $(p,2m)=(1,1)$,
the uncontracted vertex, and $(p,2m)=(2,0)$, which is zero by
symmetry;  so no renormalization of coupling constants is required.  
At $E<\m$, the size of the boson propagator becomes  $(E/\m)^2$; then
a graph size is $\m^{n-p} E^{d'(p,n,2m)}$, 
for $d'(p,n,2m)=2-m-n$, that is the
same {\it power counting} of the Thirring model. Still qualitatively, 
the limit $\m\to 0$ gives the Schwinger model; 
the limit $\m\to \io$ gives the free boson field; 
whereas replacing $G_0^{\m\n}$  with $\m^2G_0^{\m\n}$,
and taking the limit $\m\to\io$ give the Thirring model.   

In this paper we consider a fixed $\m>0$. The issue with the above
argument is that we are not  able to prove the
convergence of the perturbation theory with 
both boson and fermion fields. To overcome that
we integrate the boson field before the \rg\  analysis so
that the  fermion-boson interaction is turned into a 
fermion-fermion quartic interaction. Now the theory looks
marginal at any scale. To recover the superrenormalizability of the UV
scales  we use identities among the
Feynman  graphs: the identities of
this paper are the same as in \cite{[FM]} and \cite{[BFM3]}; as
opposed to approach in \cite{[Ma]}, \cite{[Ma2]},  they permit to take
advantage of $L_p$ inequalities, which is the key to control an
unbounded  $G_0$. This part is largely inspired to \cite{[Le]}.

Before discussing technical details, we set up some more
notations. The explicit choice of generators of the Euclidean
Clifford algebra is
$$\g^0=
 \pmatrix{0&1\cr
          1&0\cr}\;,
 \qquad
 \g^1=
 \pmatrix{0&-i\cr
          i&0\cr}\;;
$$
then we call 
$\h=(\h^-_{\xx,-}, \h^-_{\xx,+})$, $\bar\h=(\h^+_{\xx,+}, \h^+_{\xx,-})$, 
$\ps_\xx=(\ps^-_{\xx,+},\ps^-_{\xx,-})$ and 
$\bar \ps_{\xx}=(\ps^+_{\xx,-},\ps^+_{\xx,+})$ 
- note the opposite notation for the components
of the spinors $\bar \h$, $\h$  and
$\bar \ps$, $\ps$ - so that, for 
$\dpr_{\xx,\o}=i\dpr^0-\o \dpr^1$,
$$
\ps^+_{\xx,\o}\ps^-_{\xx,\o}=
\bar\ps_\xx{\g^0-i\o\g^1 \over 2}\ps_\xx\;,
\qquad
\sum_\o\ps^+_{\xx,\o}\dpr_\o\ps^-_{\xx,\o}= i
\bar\ps_\xx\g^\m\dpr^\m\ps_\xx\;.
$$
The interaction becomes 
$$
v_{\o,\o'}(\xx)={1\over 2}
\lft[i\o G_0^{10}(\xx)+i\o' G_0^{01}(\xx)+G_0^{00}(\xx)-\o\o'G_0^{11}(\xx)\rgt]\;;
$$
and since $G_0^{01}(\xx)=G_0^{10}(\xx)$, also  $v_{\o,\o'}(\xx)=v_{\o',\o}(\xx)$.
Finally, for $J_{\xx,\o}=J^0_\xx+i\o J^1_\xx$, 
the functional integral formula for the non-local Thirring model is 
\bea
e^{\WW_{l,N}(J,\h)}= 
\int\!dP_{l,N}(\psi)\;
\exp\lft\{
{\l\over 2}\sum_{\o,\o'}\int\!d\xx d\yy\; 
\psi^+_{\xx,\o}\psi^-_{\xx,\o}\;
v_{\o,\o'}(\xx-\yy)\;\psi^+_{\xx,\o'}\psi^-_{\xx,\o'}\rgt\}
\cr\cr
\exp\lft\{\sum_\o\int\!d\xx\; J_{\xx,\o}\psi^+_{\xx,\o}\psi^-_{\xx,\o}
+\sum_\o\int\!d\xx\;(\h^+_{\xx,\o}\ps^-_{\xx,\o}
+\ps^+_{\xx,\o}\h^-_{\xx,\o})\rgt\}
\eea
for  $dP_{l,N}(\ps)$ determined by the covariance:
$$
\int\!dP_{l,N}(\ps)\; \ps^-_{\xx,\o}\ps^+_{\xx,\o} 
=g^{[l,N]}_\o(\xx)=\int\!{d\kk\over (2\p)^2}\;
{e^{-i\kk\xx}\over D_\o(\kk)} \hac_{l,N}(\kk)\;,
\qquad
D(\kk)=k^0+i\o k^1\;.
$$

Finally, we stress two points. 
Firstly, we are assuming that the limit of removed cutoff in the
propagator $v_{\o,\o'}$ is already taken: 
this is not an abuse, since, otherwise, 
the estimates that will follow would be anyways uniform in the 
$l,N$ of $v_{\o,\o'}$.
Secondly, all the claims about $\WW(J,\h)$ (and the same for other
functional integrals that we are about to define), must actually be understood
in terms of  the correlations that it generates, 
i.e. for a finite number of derivatives w.r.t
the external fields, at $J=\h=0$. (In fact there would be no need to prove that
$\WW(J,\h)$ is a convergent power series of the external fields even
if we wanted to verify the Osterwalder-Schrader axioms, see \cite{[BFM1]}.)

\subsection{Correlations.}\lb{s4.1}
In evaluating $\WW_{l,N}(J,\h)$, to have bounds that are uniform in
$l$ and $N$, we have to slice the range of allowed momenta into
scales.   We use the decomposition
$$\hac_{h',h}(\kk)
=\sum_{k=h'+1}^h f_k(\kk)
$$
where $f_k(\kk)=\hac_{k-1,k}(\kk)$; in correspondence we have the
factorization of the Gaussian measure  
\bea
&&\ps=\ps^{(h')}+\ps^{(h'+1)}+\cdots+ \ps^{(h)}
\cr\cr
&&dP_{h',h}(\ps)=dP_{h'}(\ps^{(h')})dP_{h'+1}(\ps^{(h'+1)})\cdots dP_{h}(\ps^{(h)})
\eea
and
$dP_{k}(\ps^{(k)})$ is determined by the covariance
$$
g_\o^{[k]}(\xx)=\int\!{d\kk\over (2\p)^2}\
{e^{-i\kk\xx}\over D_\o(\kk)}f_k(\kk)\;.
$$
We integrate iteratively the fields with smaller and smaller
momentum. After the integration of 
$\ps=\ps^{(N)}, \ps^{(N-1)},\ldots, \ps^{(h+1)}$ we have the effective
potential on scale $h$, $\VV^{(h)}$, such that
\be\lb{ep0}\;
e^{\WW_{l,N}(J,\h)}=\int\!dP_{l,h}(\ps)\ 
e^{\VV^{(h)}(\ps, J, \h)}\;.
\ee
Consider the case $\bar\h=\h=0$.
Assume by induction that, for any scale $h= k+1$, the effective
potential is a polynomial in the fields $(J_{\zz,\o})$,
$(\ps^+_{\o,\xx})$ and $(\ps^-_{\o,\yy})$. We call kernels on scale
$h$ the coefficients of the monomials of $\VV^{(h)}$:  
for $\uz=(\zz_1,\ldots, \zz_n)$, 
$\ux=(\xx_1,\ldots, \xx_m)$,   
$\uy=(\yy_1,\ldots, \yy_m)$, 
$\uo'=(\o_1,\ldots \o_n)$ and  
$\uo=(\o_1,\ldots \o_m)$, 
\be\lb{ker}
W^{(n;2m)(h)}_{\uo',\uo}(\uz;\ux,\uy)
\defi\lft.
\prod_{j=1}^n {\dpr \over \dpr J_{\zz_i,\o'_i}}
\prod_{i=1}^{m}{\dpr \over \dpr \ps^+_{\xx_i,\o_i}}
{\dpr \over \dpr \ps^-_{\yy_i,\o_i}}
\VV^{(h)}(\ps,J,0)
\rgt|_{J=\ps=0}\;
\ee
(where the derivatives in $\dpr \ps^-_{\yy_i,\o_i}$ are taken from the
right).
To evaluate
$\VV^{(k)}(\ps,J,0)$, use the formula for the truncated expectations: 
\bea\lb{ep}
&&\VV^{(k)}(\ps,J,0)=\ln\int\!dP_{k+1}(\z)\ 
e^{\VV^{(k+1)}(\ps+\z, J, 0)}
\cr\cr
&&=\sum_{p\ge 0}{1\over p!}
E^T_{k+1}
\Big[\underbrace{\VV^{(k+1)}(\ps+\z,J,0);\cdots;
\VV^{(k+1)}(\ps+\z,J,0)}_{p {\rm \ times}}\Big] 
\eea
where $E^T_{k+1}$ is by definition the truncated expectation
w.r.t. the Gaussian random variables $(\z^\e_{\xx, \o})$ with
covariances $(g_\o^{[k+1]}(\xx))$. Accordingly, \pref{ep} gives through
\pref{ker} the kernels $W^{(n;2m)(k)}_{\uo',\uo}$. Formula \pref{ep}
gives also the well known interpretation of each kernels as sum of
Feynman graphs belonging to a given class, that is determined by the 
``external legs''.  Later on, we will  take advantage of the following
two identities on the structure of the graph expansion of the kernels.
\begin{lemma} The derivatives of the effective potential satisfy two
identities: 
\bea
&&{\dpr \VV^{(k)}\over \dpr \ps^+_{\xx,\o}}(\ps,J,0)
=J_{\xx,\o}\ps_{\xx,\o}^- 
+J_{\xx,\o}\int\! d\uu\ 
 g_\o(\xx-\uu){\dpr \VV^{(k)}\over \dpr \ps_{\uu,\o}^+}(\ps,J,0)
\lb{prc1}
\cr\cr 
&&+\l\sum_{\o,\o'}\int\!d\ww d\uu\ 
v_{\o,\o'}(\xx-\ww)g_\o(\xx-\uu)\lft[
{\dpr^2\VV^{(k)}\over\dpr J_{\ww,\o'} \dpr \ps^+_{\uu,\o}}
+{\dpr \VV^{(k)}\over\dpr J_{\ww,\o'}}
{\dpr \VV^{(k)}\over\dpr \ps^+_{\uu,\o}}\rgt](\ps,J,0)
\cr\cr 
&&+\l\sum_{\o,\o'}\int\!d\ww\ 
v_{\o,\o'}(\xx-\ww)
\ps_{\xx,\o}^- {\dpr \VV^{(k)}\over\dpr J_{\ww,\o'}}(\ps,J,0)\;,
\\
\cr\cr
&&{\dpr \VV^{(k)}\over \dpr J_{\xx,\o}}(\ps,J,\h)
=\ps^+_{\xx,\o}\ps^-_{\xx,\o}+\int\!d\uu\ 
g_\o(\xx-\uu)\lft[
\ps^+_{\xx,\o}{\dpr \VV^{(k)}\over \dpr \ps^+_{\uu,\o}}-
{\dpr \VV^{(k)}\over \dpr \ps^-_{\uu,\o}}\ps^-_{\xx,\o}
\rgt](\ps,J,\h)
\cr\cr
&&+\int\!d\uu d\uu'\ 
g_\o(\xx-\uu)g_\o(\xx-\uu')
\lft[
{\dpr^2 \VV^{(k)}\over \dpr \ps^+_{\uu,\o}\dpr \ps^-_{\uu',\o}}+
{\dpr\VV^{(k)}\over \dpr \ps^+_{\uu,\o}}
{\dpr\VV^{(k)}\over\dpr \ps^-_{\uu',\o}}\rgt](\ps,J,\h)\;.
\lb{prc2}\eea
\end{lemma}
These identities are clear from graphical interpretation of 
the multiscale integration; in appendix \ref{a1} we will prove
them from the definition of $\VV^{(k)}$ and $\VV$. 

We introduce the following $L_1$ norm
\bea\lb{norm}
\|W^{(n;2m)(k)}_{\uo',\uo}\|=
\int\! d\ux d\uy d\uz_2\;
\lft|W^{(n;2m)(k)}_{\uo',\uo}
(\uz;\ux,\uy)\rgt|
\eea
for $d\uz_2=(\zz_2,\ldots,\zz_m)$; namely in \pref{norm} we are
integrating all but one variable; by translation invariance,
the norm does not depend upon $\zz_1$. Since $W^{(n;2m)(k)}_{\uo',\uo}$ may
contain delta-distributions, we extend the definition of $L_1$ norm by
considering them as positive functions.

Let $\m^2=\g^{2M}$. We will use the following straightforward bounds, 
for $c,c_p,c', B, B_p>1$:
\bea\lb{dec}
&&\|g_\o^{(h)}\|_{L_\io}\defi \sup_{\xx} |g_\o^{(h)}(\xx)|
\le c\g^{h}\;,
\cr\cr
&&\|g_\o^{(h)}\|_{L_p}\defi
\lft[\int\!d\xx\;|g_\o^{(h)}(\xx)|^p\rgt]^{1/p}\le c_p
\g^{\lft(1-{2\over p}\rgt)h}
\cr\cr
&&
\|g_\o^{(h)}\|_{L_1(w)}\defi \int\!d\xx\;|x_j||g_\o^{(h)}(\xx)|\le 
c' \g^{-2h}
\eea
and, since $\a\kk^2+\m^2-\s\ge\a_0[\kk^2+\a_0^{-1}(\m^2-\s)]$,
uniformly in $\a\ge \a_0$
\bea\lb{dec2}
&&\|v_{\o,\o'}\|_{L_p}\defi \lft[\int\!d\xx |v_{\o,\o'}(\xx)|^p\rgt]^{1/p}\le B_p
\g^{2\lft(1-{1\over p}\rgt)M}
\cr\cr
&&\|\dpr_j
v_{\o,\o'}\|_{L_1}\defi\int\!d\xx\;|(\dpr_jv_{\o,\o'})(\xx)|
\le B \g^{M}
\;.
\eea
Let's consider separately the two different regimes: the UV
one, that corresponds to the scales $k:M\le k\le N$, and the IR, for
$k:l\le k\le M-1$.

Define
\bea
&&w^{(1;2)}_{\o',\o}(\zz,\xx,\yy)=\d(\zz-\xx)\d(\zz-\yy)\d_{\o,\o'}
\cr\cr
&&w^{(0;4)}_{\o',\o}(\xx,\yy,\uu,\vv)=\d(\xx-\yy)v_{\o',\o}(\xx-\uu)\d(\uu-\vv)
\eea
\insertplot{420}{35}
{\ins{112pt}{9pt}{$\zz$}
\ins{125pt}{37pt}{$\o$}
\ins{125pt}{0pt}{$\o$}

\ins{218pt}{9pt}{$\xx$}
\ins{208pt}{37pt}{$\o$}
\ins{208pt}{0pt}{$\o$}
\ins{247pt}{9pt}{$\uu$}
\ins{254pt}{40pt}{$\o'$}
\ins{254pt}{2pt}{$\o'$}

}%
{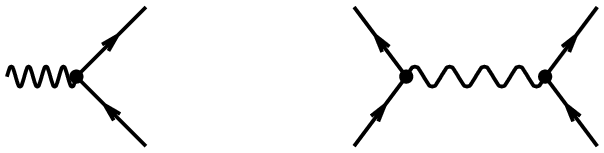}{\lb{p0} Graphical representation for $w^{(1;2)}_{\o',\o}$ and 
$w^{(0;4)}_{\o',\o}$}{0}
and note that at $h=N$ we have $W^{(0;2)(N)}_{\o}(\xx,\yy)=0$, 
$W^{(1;2)(N)}_{\o',\o}(\zz,\xx,\yy)=w^{(1;2)}_{\o',\o}(\zz,\xx,\yy)$
and $W^{(0;4)(N)}_{\o',\o}(\xx,\yy,\uu,\vv)
=\l w^{(0;4)}_{\o',\o}(\xx,\yy,\uu,\vv)$; all the other
$W^{(n;2m)(N)}_{\uo',\uo}$ are zero. 

\begin{theorem}\lb{t3.2}
For $|\l|$ small enough, there exist constants $C_0,C>1$ 
such that, for $M\le h\le N$ 
\bea\lb{hb}
&&\|W^{(0;2)(h)}_{\o}\|
\le C|\l|\g^h
\;,
\cr\cr
&&\|W^{(1;2)(h)}_{\o',\o}-w^{(1;2)}_{\o',\o}\|
\le C|\l|
\;,
\cr\cr
&&\|W^{(0;4)(h)}_{\o',\o}-\l w^{(0;4)}_{\o',\o}\|
\le C|\l|
\;;
\eea
and, for any other $(n;2m)$  
\be\lb{hb2}
\|W^{(n;2m)(h)}_{\uo',\uo}\|
\le C_0^{n+d_{n,2m}}(C|\l|)^{d_{n,2m}}\g^{h(2-n-m)}
\ee
where $d_{0,2}=1$, $d_{n,0}=0$, otherwise $d_{n,2m}=m-1$.
\end{theorem} 
The point in the bounds is that $C$, $C_0$ are $N-h$ independent. 
The proof of \pref{hb2} for $h=k$, assumed iteratively \pref{hb} and
\pref{hb2} for $h\ge k+1$, is standard. We shall focus, therefore, on
\pref{hb} that improves \pref{hb2} in the cases of 
{\it marginal and relevant graphs}, i.e.
$(n;2m)=(0;2),(0;4), (1;2)$.

{\bf\0Proof.}
To shorten the notation, in this proof we define  
$\z\defi \psi^{(k+1)}+\psi^{(k+2)}+\cdots+\psi^{(N)}$ and 
$g_\o\defi g_\o^{[k+1,N]}$. The derivatives in $\ps^-$, $\h^-$ and
$\z^-$ are taken from the right.
The proof is for  $C$ large enough with respect to 
$C_0,c, c_p, c', B, B_p$.
\\

{\it\01. Improved bound for $(0;2)$.}  By symmetry we
have $W^{(1;0)(k)}_{-\o}(\ww)\=0$; hence
from \pref{prc1} and \pref{prc2}
we expand the two-points kernel as in Fig.\ref{p1}
\insertplot{420}{40}
{\ins{104pt}{23pt}{$\o$}
\ins{161pt}{23pt}{$\o$}
\ins{112pt}{12pt}{$\xx$}
\ins{153pt}{12pt}{$\yy$}

\ins{178pt}{16pt}{$=$}

\ins{198pt}{23pt}{$\o$}
\ins{278pt}{23pt}{$\o$}
\ins{202pt}{12pt}{$\xx$}
\ins{226pt}{1pt}{$\ww'$}
\ins{228pt}{42pt}{$\ww$}
\ins{272pt}{12pt}{$\yy$}
}%
{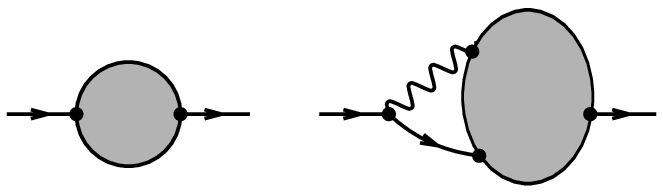}{\lb{p1} Decomposition of the class of graphs  $W_\o^{(0;2)(k)}$.}{0}
\bea\lb{111}
W^{(0;2)(k)}_{\o}(\xx,\yy)=
\l \sum_{\o'}\int\!d\ww d\ww'\; 
v_{\o,\o'}(\xx-\ww) g_\o(\xx-\ww')
W^{(1;2)(k)}_{\o';\o}(\ww;\ww',\yy)\;,
\eea
so that, from $\|w_{\o';\o}^{(1;2)}\|\le 1$ and from \pref{hb2} for 
$(n;2m)=(1;2)$, we obtain, for $C$ large enough,
\bea
\lb{111b}
\|W^{(0;2)(k)}_{\o}\|
\le |\l|(1+C_0)  
\sum_{\o'} \|v_{\o,\o'}\|_{L_3} \sum_{j=k}^N \|g^{(j)}_\o\|_{L_{3/2}}
\le {C\over r_1}|\l|\g^k\g^{-{4\over 3}(k-M)}\;,
\eea
that proves the first of \pref{hb}. The factor $r_1>1$ will be useful
for later. 
\\

{\it\02. Improved bound for $(1;2)$.} By \pref{prc1}, the kernel 
$W^{(1;2)(k)}_{\o';\o}$ can be rewritten as in Fig.\ref{p2}.
\insertplot{420}{140}
{\ins{46pt}{108pt}{$\o$}
\ins{52pt}{97pt}{$\xx$}
\ins{99pt}{108pt}{$\o$}
\ins{93pt}{97pt}{$\yy$}
\ins{62pt}{129pt}{$\o'$}
\ins{81pt}{121pt}{$\zz$}

\ins{130pt}{102pt}{$=$}

\ins{157pt}{108pt}{$\o$}
\ins{163pt}{97pt}{$\xx$}
\ins{190pt}{86pt}{$\uu$}
\ins{234pt}{98pt}{$\o$}
\ins{225pt}{86pt}{$\yy$}
\ins{190pt}{130pt}{$\ww$}
\ins{234pt}{133pt}{$\o'$}
\ins{225pt}{130pt}{$\zz$}
\ins{205pt}{143pt}{(a)}

\ins{50pt}{43pt}{$+$}

\ins{59pt}{19pt}{$\o$}
\ins{64pt}{6pt}{$\xx=\yy$}
\ins{79pt}{26pt}{$\ww$}
\ins{62pt}{67pt}{$\o'$}
\ins{81pt}{57pt}{$\zz$}
\ins{95pt}{60pt}{(b)}

\ins{117pt}{43pt}{$+$}

\ins{141pt}{18pt}{$\o$}
\ins{150pt}{7pt}{$\xx$}
\ins{173pt}{7pt}{$\uu$}
\ins{211pt}{18pt}{$\o$}
\ins{205pt}{7pt}{$\yy$}
\ins{159pt}{26pt}{$\ww$}
\ins{143pt}{67pt}{$\o'$}
\ins{161pt}{57pt}{$\zz$}
\ins{205pt}{60pt}{(c)}

\ins{230pt}{43pt}{$+\ \d_{\o',\o}$}

\ins{273pt}{27pt}{$\xx=\zz$}
\ins{270pt}{38pt}{$\o$}
\ins{303pt}{27pt}{$\uu$}
\ins{343pt}{38pt}{$\o$}
\ins{337pt}{27pt}{$\yy$}
\ins{295pt}{60pt}{(d)}
}%
{p2}{\lb{p2} Decomposition of the class of graphs  $W^{(1;2)(k)}_{\o';\o}$. The
darker bubble is for $W^{(1;2)(k)}_{\o';\o}-w^{(1;2)}_{\o';\o}$}{0}
Graph (a) in Fig.\ref{p2} is  given by:
\bea
W^{(1;2)(k)}_{(a)\o';\o}(\zz;\xx,\yy)
\defi
\l\sum_{\o''}\int\! d\ww d\uu\
v_{\o,\o''}(\xx-\ww)g_\o(\xx-\uu)W^{(2;2)(k)}_{\o',\o'';\o}(\zz,\ww;\uu,\yy) 
\eea
From \pref{hb2} for $(n;2m)=(2;2)$, we obtain
\bea
\|W^{(1;2)(k)}_{(a);\o';\o}\|
\le |\l|C_0^2\g^{-k} \sum_{\o''}\|v_{\o,\o''}\|_{L_3}
\sum_{j=k}^N \|g^{(j)}_\o\|_{L_{3/2}}
\le {C\over 4r_2}|\l|\g^{-{4\over3} (k-M)}
\eea
where a large enough constant  $r_2>1$ will be used later.
For graphs (c) and (d) we use the just improved bound for $
W^{(0;2)(k)}_{\o}$: for instance, 
 graph (d) is given by
\bea
W^{(1;2)(k)}_{(d)\o';\o}(\zz;\xx,\yy)
\defi
\d_{\o,\o'}
\d(\xx-\zz)\int\! d\uu\ 
g_\o(\xx-\uu) W^{(0;2)(k)}_{\o}(\uu,\yy)
\eea
and using \pref{111b} for a $r_1$ large enough to compensate other
constants, we get 
\bea
\|W^{(1;2)(k)}_{(d)\o';\o}(\zz;\xx,\yy)\|
\le
\| W^{(0;2)(k)}_\o\|
\sum_{j=k}^N \|g^{(j)}_\o\|_{L_1}\le 
{C \over 4r_2}|\l|\g^{-{4\over3}(k-M)}\;.
\eea
In order to obtain an improved bound also for the 
graphs (b) of Fig.\ref{p2}, we need to further 
expand $W^{(2;0)(k)}_{\o';-\o}$. Using \pref{prc2}, we find
\bea\lb{63}
W^{(2;0)(k)}_{\o',-\o}(\zz,\ww)
 =
\int\! d\uu'd\uu\ 
g_{\o}(\ww-\uu)g_{\o}(\ww-\uu')
W^{(1;2)(k)}_{\o';-\o}(\zz;\uu',\uu)
\eea
and then, replacing the expansion 
for $W^{(1;2)(k)}_{\o';-\o}(\zz;\uu',\uu)$
in the graph \pref{63} we find
for (b) what is depicted in Fig.\ref{p3}.
Graphs (b4) and (b5) have been obtained also using the expansion \pref{111}.
\insertplot{420}{155}
{\ins{51pt}{142pt}{$\o$}
\ins{51pt}{127pt}{$\xx$}
\ins{73pt}{125pt}{$\ww$}
\ins{100pt}{153pt}{$\uu'$}
\ins{100pt}{114pt}{$\uu$}
\ins{134pt}{145pt}{$\o'$}
\ins{128pt}{125pt}{$\zz$}
\ins{63pt}{153pt}{(b)}

\ins{166pt}{132pt}{$=$}

\ins{186pt}{142pt}{$\o$}
\ins{186pt}{126pt}{$\xx$}
\ins{208pt}{124pt}{$\ww$}
\ins{220pt}{146pt}{$\uu'$}
\ins{246pt}{128pt}{$\zz'$}
\ins{267pt}{116pt}{$\uu$}
\ins{267pt}{156pt}{$\ww'$}
\ins{285pt}{145pt}{$\o'$}
\ins{277pt}{124pt}{$\zz$}
\ins{198pt}{159pt}{(b1)}

\ins{25pt}{84pt}{$+$}

\ins{51pt}{92pt}{$\o$}
\ins{51pt}{75pt}{$\xx$}
\ins{73pt}{73pt}{$\ww$}
\ins{130pt}{94pt}{$\o'$}
\ins{122pt}{73pt}{$\zz$}
\ins{63pt}{103pt}{(b2)}

\ins{167pt}{84pt}{$+$}

\ins{186pt}{92pt}{$\o$}
\ins{186pt}{75pt}{$\xx$}
\ins{209pt}{73pt}{$\ww$}
\ins{257pt}{73pt}{$\uu$}
\ins{267pt}{76pt}{$\zz'$}
\ins{306pt}{94pt}{$\o'$}
\ins{300pt}{73pt}{$\zz$}
\ins{198pt}{103pt}{(b3)}

\ins{25pt}{35pt}{$+$}

\ins{51pt}{43pt}{$\o$}
\ins{51pt}{25pt}{$\xx$}
\ins{70pt}{23pt}{$\ww$}
\ins{150pt}{45pt}{$\o'$}
\ins{142pt}{23pt}{$\zz$}
\ins{72pt}{14pt}{$\ww'$}
\ins{134pt}{14pt}{$\zz'$}
\ins{104pt}{2pt}{$\vv'$}
\ins{104pt}{32pt}{$\vv$}
\ins{63pt}{53pt}{(b4)}

\ins{167pt}{33pt}{$+$}

\ins{186pt}{43pt}{$\o$}
\ins{186pt}{25pt}{$\xx$}
\ins{205pt}{23pt}{$\ww$}
\ins{277pt}{23pt}{$\uu$}
\ins{208pt}{14pt}{$\ww'$}
\ins{270pt}{14pt}{$\zz'$}
\ins{286pt}{26pt}{$\uu'$}
\ins{240pt}{2pt}{$\vv'$}
\ins{240pt}{32pt}{$\vv$}
\ins{330pt}{43pt}{$\o'$}
\ins{324pt}{23pt}{$\zz$}
\ins{198pt}{53pt}{(b5)}

}%
{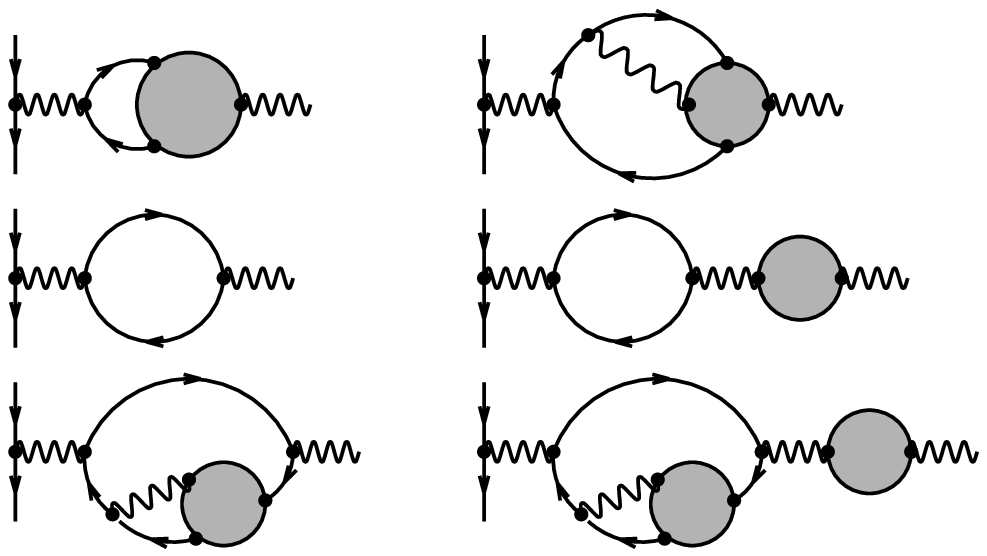}{\lb{p3}Further decomposition of the class of graphs (b) in Fig.\ref{p2} }{0}
A bound for (b2) and (b3) can be found in the same way. Consider, for
instance, the expression for (b2):
\bea\lb{w12}
W^{(1;2)(k)}_{(b2)\o';\o}(\zz;\xx,\yy)
\defi
\l \d(\xx-\yy)\int\!d\ww\ 
v_{\o',\o}(\xx-\ww)
g^2_{\o'}(\ww-\zz)
\eea
We want to use the cancellation $\int\!  d\uu\ g^2_{\o}(\uu)=0$ that
is a consequence of the symmetry under rotation of the model. In order
to to that, expand
\be \lb{idb}
v_{\o,\o}(\xx-\ww)=v_{\o,\o'}(\xx-\zz)+
\sum_{j=0,1} (z_j-w_j) \int_0^1\!\!d \t\ 
\big(\dpr_j v_{\o,\o'}\big)\big(\xx-\zz+\t(\zz-\ww)\big)
\ee
and plug \pref{idb} into \pref{w12}: one term in zero; the other can
be bounded as follows: 
\bea
\|W^{(1;2)(k)}_{(b2)\o';\o}\|
\le
4|\l| \|\dpr_j v_{\o,\o'}\|_{L_1}
\sum_{i=k}^N\sum_{j=k}^i
\|g^{(j)}_{\o'}\|_{L_\io}
\|g^{(i)}_{\o}\|_{L_1(w)}
\le |\l| {C\over 20r_2}\g^{-(k-M)}
\eea
Now consider (b1)
\bea
W^{(1;2)(k)}_{(b1)\o';\o}(\zz;\xx,\yy)
\defi
\l\d(\xx-\yy)\sum_{\s,\s'}\int\! d\ww d\uu' d\zz'\ 
v_{\o,\s}(\xx-\ww) v_{\s,\s'}(\uu'-\zz')\cdot
\cr\cr
\cdot\int\! d\uu  d\ww'\ g_{\s}(\ww-\uu)g_{\s}(\ww-\uu')
g_{\s}(\uu'-\ww')
W^{(2;2)(k)}_{\o',\s';\s}(\zz,\zz';\ww',\uu)\;;
\eea
therefore
\bea\lb{27}
\|W^{(1;2)(k)}_{(b1)\o';\o}\|
\le 
|\l|\|v_{\o,\s}\|_{L_1}
\int\! d\ww'\; d\uu\;  d\zz'\ \ 
|W^{(2;2)(k)}_{\o',\s';\s}(0,\zz';\ww',\uu)|
\cdot\cr\cr\cdot
\int\!d\uu'\; d\ww\;\;|v_{\s,\s'}(\uu'-\zz')
g_{\s}(\ww-\uu)g_{\s}(\ww-\uu') g_{\s}(\uu'-\ww')|\;.
\eea
We have to find  a bound for the second line that is uniform in
$N-k$. For that, it is convenient to 
decompose the three fermion propagators into scales,
$\sum_{j,q,p=k}^N g^{(j)}_{\s}g^{(q)}_{\s} g^{(p)}_{\s}$
and then, for each realization of $j,q,p$, we take the $\|\cdot\|_{L_1}$ on the 
fermion propagator with lowest scale. This is always possible:
\\
for $p\le q,j$
\bea
\int\!d\ww\; d\uu'\;|v_{\s,\s'}(\uu'-\zz')
g^{(j)}_{\s}(\uu'-\ww) g^{(q)}_{\s}(\uu'-\ww')g^{(p)}_{\s}(\ww-\uu)|
\cr\cr
=\int\!d\uu'\;|v_{\s,\s'}(\uu'-\zz')
 g^{(q)}_{\s}(\uu'-\ww')| \int\!d\ww\;
|g^{(j)}_{\s}(\uu'-\ww)g^{(p)}_{\s}(\ww-\uu)|
\cr\cr
\le \|v_{\s,\s'}\|_{L_3}\|g^{(q)}_{\o}\|_{L_{3/2}}
\|g^{(j)}_{\s}\|_{L_1}\|g^{(p)}_{\s}\|_{L_\io}
\le C_3 \g^{{4\over 3}M}\g^{-{q\over 3}} \g^{-j} \g^{p} 
\eea
for $q\le p,j$
\bea
\int\!d\ww\; d\uu'\;|v_{\s,\s'}(\uu'-\zz')
g^{(j)}_{\s}(\uu'-\ww) g^{(q)}_{\s}(\uu'-\ww')g^{(p)}_{\s}(\ww-\uu)|
\cr\cr
=\int\!d\ww\;
|g^{(p)}_{\s}(\ww-\uu)|\int\!d\uu'\;|v_{\s,\s'}(\uu'-\zz')
 g^{(q)}_{\s}(\uu'-\ww')g^{(j)}_{\s}(\uu'-\ww)| 
\cr\cr
\le \|g^{(p)}_{\s}\|_{L_1}
 \|v_{\s,\s'}\|_{L_3}\|g^{(j)}_{\s}\|_{L_{3/2}}
 \|g^{(q)}_{\s}\|_{L_\io}
\le C_3\g^{{4\over 3}M} \g^{-p} \g^{-{j\over 3}} \g^{q} 
\eea
and finally, for $j\le p,q$
\bea
\int\!d\ww\; d\uu'\;|v_{\s,\s'}(\uu'-\zz')
g^{(j)}_{\s}(\uu'-\ww) g^{(q)}_{\s}(\uu'-\ww')g^{(p)}_{\s}(\ww-\uu)|
\cr\cr
=\int\!d\ww\;
|g^{(p)}_{\s}(\ww-\uu)|\int\!d\uu'\;|v_{\s,\s'}(\uu'-\zz')
 g^{(q)}_{\s}(\uu'-\ww')g^{(j)}_{\s}(\uu'-\ww)| 
\cr\cr
\le \|g^{(p)}_{\s}\|_{L_1}
 \|v_{\s,\s'}\|_{L_3}\|g^{(q)}_{\s}\|_{L_{3/2}}
 \|g^{(j)}_{\s}\|_{L_\io}
\le C_3 \g^{{4\over 3}M}\g^{-p} \g^{-{q\over 3}} \g^{j} 
\eea
so that, summing over the scales $q,p,j$, we obtain
\bea\lb{prc10}
\int\!d\ww\; d\uu'\;|v_{\s,\s'}(\uu'-\zz')
g_{\s}(\uu'-\ww) g_{\s}(\uu'-\ww')g_{\s}(\ww-\uu)|
\le C_4\g^{{4\over 3}M} \g^{-{k\over 3}}
\eea
From \pref{prc10} and \pref{27}, we obtain:
\bea
\|W^{(1;2)(k)}_{(b1)\o';\o}\| 
\le {C\over 20}
|\l| \g^{-{4\over 3}(k-M)}
\eea
Finally, the latter argument applies also to the bounds  of (b4) and
(b5).
For instance,  the expression for (b4) is
\bea
W^{(1;2)(k)}_{(b4)\o';\o}(\zz;\xx,\yy)
\defi
\d(\xx-\yy)
\l^2\int\! d\zz' d\ww\  
v_{\o,\o'}(\xx-\ww) g_{\o'}(\ww-\zz)
\cdot\cr \cr\cdot
\sum_\s\int\! d\ww'd\uu'd\uu\ 
g_{\o'}(\ww-\ww')g_{\o'}(\ww'-\uu)
v_{\o',\s}(\ww'-\uu')
\cdot\cr \cr\cdot
W^{(1;2)(k)}_{\s;\o'}(\uu';\uu,\zz') g_{\o'}(\zz'-\zz)
\eea
Hence, the bound for such a kernel is:
\bea
\|W^{(1;2)(k)}_{(b4)\s';\s}\|
\le
2|\l|^2\|v_{\o,\o'}\|_{L_1} 
\int\! d\zz'\;d\uu'\; d\uu\;
 |W^{(1;2)(k)}_{-\o;\o}(\uu';\uu,\zz') g_{\o}(\zz')|
\cr\cr\cdot
\int\! d\ww\;  d\ww'\; |g_{\o}(\ww-\zz)
g_{\o}(\ww-\ww')g_{\o}(\ww'-\uu)
v_{\o',\s}(\ww'-\uu')|
\eea
that by \pref{prc10} becomes
\bea
\|W^{(1;2)(k)}_{(b4);\o';\o}\|\le {C\over20r_2} |\l| \g^{-{4\over 3}(k-M)}
\eea
Therefore we have proved 
\be\lb{p12}
\|W^{(1;2)(k)}_{\o;\o'}-w^{(1;2)}_{\o;\o'}\|\le {C\over r_2}
|\l|\g^{-{4\over 3}k}
\ee
that is the second \pref{hb}.  
\\

{\it\03. Improved bound for $(0;4)$.}
By \pref{prc1} we obtain the identity in Fig.\ref{p4}.
\insertplot{420}{180}
{\ins{70pt}{164pt}{$\o$}
\ins{76pt}{153pt}{$\xx$}
\ins{125pt}{148pt}{$\o$}
\ins{107pt}{140pt}{$\yy$}
\ins{125pt}{180pt}{$\o'$}
\ins{111pt}{180pt}{$\yy'$}
\ins{125pt}{167pt}{$\o'$}
\ins{116pt}{154pt}{$\xx'$}

\ins{165pt}{157pt}{$=$}

\ins{190pt}{154pt}{$\o$}
\ins{188pt}{140pt}{$\xx=\yy$}
\ins{224pt}{143pt}{$\o$}
\ins{221pt}{168pt}{$\ww$}
\ins{267pt}{185pt}{$\o'$}
\ins{257pt}{183pt}{$\yy'$}
\ins{267pt}{161pt}{$\o'$}
\ins{257pt}{148pt}{$\xx'$}
\ins{190pt}{175pt}{(a)}

\ins{24pt}{95pt}{$+$}

\ins{54pt}{98pt}{$\o$}
\ins{60pt}{85pt}{$\xx$}
\ins{85pt}{75pt}{$\uu$}
\ins{80pt}{112pt}{$\ww$}
\ins{127pt}{85pt}{$\o$}
\ins{117pt}{75pt}{$\yy$}
\ins{127pt}{105pt}{$\o'$}
\ins{111pt}{95pt}{$\xx'$}
\ins{127pt}{127pt}{$\o'$}
\ins{112pt}{126pt}{$\yy'$}
\ins{60pt}{123pt}{(b)}

\ins{165pt}{93pt}{$+$}

\ins{184pt}{98pt}{$\o$}
\ins{194pt}{86pt}{$\xx$}
\ins{221pt}{120pt}{$\ww$}
\ins{221pt}{74pt}{$\uu$}
\ins{269pt}{124pt}{$\o'$}
\ins{257pt}{122pt}{$\yy'$}
\ins{269pt}{107pt}{$\o'$}
\ins{260pt}{107pt}{$\xx'$}
\ins{269pt}{84pt}{$\o$}
\ins{257pt}{78pt}{$\yy$}
\ins{190pt}{123pt}{(c)}

\ins{23pt}{35pt}{$+\d_{\o,\o'}$}

\ins{55pt}{28pt}{$\o$}
\ins{85pt}{18pt}{$\o$}
\ins{48pt}{16pt}{$\xx=\yy'$}
\ins{86pt}{42pt}{$\ww$}
\ins{128pt}{55pt}{$\o$}
\ins{111pt}{54pt}{$\yy$}
\ins{128pt}{32pt}{$\o$}
\ins{112pt}{25pt}{$\xx'$}
\ins{60pt}{55pt}{(d)}

\ins{165pt}{35pt}{$+\d_{\o,\o'}$}

\ins{195pt}{28pt}{$\o$}
\ins{205pt}{17pt}{$\xx$}
\ins{228pt}{44pt}{$\ww$}
\ins{229pt}{6pt}{$\uu$}
\ins{268pt}{55pt}{$\o$}
\ins{255pt}{54pt}{$\yy$}
\ins{268pt}{32pt}{$\o$}
\ins{253pt}{25pt}{$\xx'$}
\ins{268pt}{15pt}{$\o$}
\ins{258pt}{5pt}{$\yy'$}
\ins{190pt}{55pt}{(e)}
}%
{p4}{\lb{p4}Decomposition of the class of graphs $W^{(0;4)(k)}$. 
The darker bubbles represent
$W^{(0;4)(k)}_{\o,\o'}-\l w^{(0;4)}_{\o,\o'}$ and 
$W^{(1;2)(k)}_{\o;\o'}-w^{(1;2)}_{\o;\o'}$. }{0}
\\
The bound for the sum of the graphs (a), (b), (d), and (e), all
together, 
can be easily obtained from \pref{p12}: for $r_2$ large enough
\bea
|\l|\|v_{\o,\o'}\|_{L_1} 
\|W^{(1;2)(k)}_{\o;\o'}-w^{(1;2)}_{\o;\o'}\|
\lft(1+\|g_\o\|_{L_1} \|W^{(0;2)(k)}_{\o}\|\rgt)
\le |\l|{C\over 2}\g^{-{4\over 3}(k-M)}\;.
\eea
To bound (c), use \pref{hb2} for $(n;2m)=(1;4)$, to get, for 
$|\l|$ small enough, 
\bea
\|W^{(0;4)(k)}_{(a),\o;\o,\o'}\|
\le
|\l| \|v_{\o,\o'}\|_{L_3}C_0^2 C|\l|\|g_\o\|_{L_{3/2}}
\le {C\over 2} |\l| \g^{-{4\over 3}(k-M)}
\eea
The proof of theorem \ref{t3.2} is complete. \qed

The analysis for $\h^+=\h^-\neq0$ is not different, because the
monomials in  the effective potential that are proportional to at
least one field $\h^+$ or $\h^-$ multiply a kernel that doesn't need
any power counting improvement. This is important for pointwise
estimations on correlations: w.r.t. the $L_1$ bounds of the
kernels the pointwise estimates have some missing integrations; 
but they never involve 
$W^{(0;2)}$, $W^{(1;2)}$, $W^{(0;4)}$, where, as we showed,
missing integrations would spoil the bound.

On scales $k\le M$ the above arguments do not give a power counting
improvement: the factors of type $\g^{-\th (k-M)}$ that we have seen in the
estimates so far  are unbounded. Indeed at IR regime 
the interaction $v_{\o,\o'}$ is
effectively local, namely the system is effectively like a Thirring
model. The RG approach to use is the one in \cite{[BFM1]}: with respect to
that paper, the UV scale now is replace by $M$;  and to the
interaction, that there is purely quartic, now has a further term, that is
irrelevant, generated by the integrations of the scales $[M,N]$.  

We do not repeat all the details of \cite{[BFM1]}. We just stress
that $W^{(0;2)}$, $W^{(1;2)}$, $W^{(0;4)}$  must be {\it localized} to extract the 
relevant part of the interaction. 
In this way,  $W^{(0;2)}$ causes the flow of
the field renormalization, which is responsible of the anomalous
dimension of the large distance decay of fermion
correlations. $W^{(1;2)}$ causes the flow of
the current renormalization, that is responsible for the anomalous
dimension of current correlations. Finally $W^{(0;4)}$
changes scale by scale the effective coupling of the
quartic self-interaction: the related flow stays bounded thanks to
the vanishing of the beta function, asymptotically for $M-k\to \io$;
this crucial property is not modified by the additional irrelevant
interaction on scale $M$, see \cite{[BFM1]}. 

The final result of the multiscale integration is that there exists
the   limit of removed cutoff of
the two point correlations; and for
large $|\xx-\yy|$,
$$
{\dpr^2 \WW\over \dpr\h^+_{\xx,\o} \dpr\h^-_{\yy,\o}}(0,0)
\sim{C\over|\xx-\yy|^{1+\h}}\;,
\qquad
{\dpr^2 \WW\over \dpr J_{\xx,\o} \dpr J_{\yy,\o}}(0,0)
\sim{C\over|\xx-\yy|^{2+y}}\;.
$$
At this stage we do not know yet the explicit
expressions \pref{sol1}, \pref{sol2} of the two above correlation; nor we know the formula for
$\h$ and the fact that $y=0$. We only have a convergent power series
for them.  Explicit evaluations come from \wi\  and \sde.

\subsection{Ward Identities}\lb{s4.2}
By definition of $J_{\xx,\o}$,  we have ${\dpr\over \dpr J_{\xx,\o}}={1\over 2}
\lft[{\dpr\over \dpr J^0_{\xx}}-i\o{\dpr\over \dpr J^1_{\xx}}\rgt]$.
The analysis of this section will be done in Fourier transform: these
are the conventions
$$
\ps^\e_{\xx,\o}=\int{d\kk\over (2\p)^2}\;
e^{i\e\kk\xx}\; \hp^\e_{\kk,\o}\;,
\qquad
\h^\e_{\xx,\o}=\int{d\kk\over (2\p)^2}\;
e^{i\e\kk\xx}\; \hh^\e_{\kk,\o}\;,
$$
while  $J_{\xx,\o}$ and $v_{\o,\o'}(\xx)$ follow the same
convention of $\ps^-_{\xx,\o}$ and $\ps^+_{\xx,\o}$, respectively.
Setting
$$
\n_{\o,\o'}(\pp)={\l\over 4\p}\hat v_{\o,\o'}(\pp)\;,\qquad
B_{\o}(\pp)= \int\!{d\kk\over (2\p)^2}
 \lft[\hh^+_{\kk+\pp,\o}
 {\partial \WW\over \partial \hh^+_{\kk,\o}}-
 {\partial \WW\over \partial \hh^-_{\kk+\pp,\o}}
 \hh^-_{\kk,\o}\rgt]\;,
$$
we want to prove that the correlation functions satisfy, in the limit
of removed cutoffs,  identities generated by the following equation
for $\WW$
\be\lb{wi}
D_{\o}(\pp){\partial \WW\over \partial
\hJ_{\pp,\o}}
-D_{-\o}(\pp)\sum_{\o'}\n_{\o,\o'}(\pp)
{\partial \WW\over\partial \hJ_{\pp,\o'}}-
 {1\over 4\p}D_{-\o}(\pp)\hJ_{-\pp,\o}=B_{\pp,\o}(J,\h)\;.
\ee
Since $\n_{\o,\o'}(\pp)=\n_{\o',\o}(\pp)$, 
summing \pref{wi} over $\o$ we find
%
$$\sum_{\o,\o'} D_{\o}(\pp)\lft[\d_{\o,\o'}-\n_{-\o,\o'}(\pp)\rgt]
{\partial \WW\over \partial \hJ_{\pp,\o'}}-
{1\over 4\p}\sum_{\o} D_{\o}(\pp)\hJ_{-\pp,-\o}=\sum_\o B_{\pp,\o}(J,\h)\;,
$$
that is \pref{vec} for $\n=\l\Th={\l\over 4\p}$. 
Whereas multiplying \pref{wi} times $\o$ and summing over
$\o$ we find
%
$$
\sum_{\o,\o'} \o D_{\o}(\pp)\lft[\d_{\o,\o'}+\n_{-\o,\o'}(\pp)\rgt]
{\partial \WW\over \partial \hJ_{\pp,\o'}}+
{1\over 4\p}\sum_{\o}\o D_{\o}(\pp)\hJ_{-\pp,-\o}=\sum_\o \o B_{\pp,\o}(J,\h)\;,
$$
that is \pref{ax} for $\n_5=\l\Th_5=-{\l\over 4\p}$.
In order to prove \pref{wi}, use a general combination of the vector and axial vector
transformations: for a real  $\ha_{\pp,\o}$  (with the same Fourier
transform convention as $\hJ_{\pp,\o}$) and transform the fields 
in  $\WW_{l,N}(\h,J)$ as follows
$$
\hp^\e_{\kk,\o}\to 
\hp^\e_{\kk,\o}+
\e\int\!{d \pp\over (2\p)^2}\ha_{\pp,\o}\hp^\e_{\kk+\e\pp,\o}\;;
$$
that gives the identity
\be\lb{12}
D_\o(\pp){\partial \WW_{l,N} \over \partial
\hJ_{\pp,\o}}(J,\h) = B_{\pp,\o}(J,\h)+R_{\o,l,N}(\pp;J,\h)
\ee
where $R_{\o,l,N}(\pp;J,\h)$ is a remainder w.r.t. the formal \wi: 
the presence in the free measure of the cutoff function
$\chi_{l,N}(\kk)$ explicitly breaks the vector and axial-vector
symmetries. To study $R_{\o,l,N}(\pp;J,\h)$ we introduce a new
functional integral,
\bea
e^{\HH_{l,N}(\a,J,\h)}=\int\! dP_{l,N}(\ps)\; 
e^{\VV(\ps,\h,J)+\AA_0(\ps,\a)-\AA_{-}(\ps,\a)}
 \nn\eea
where
\bea
&&\AA_0 (\a,\ps)
=
\sum_{\o=\pm}\int\! {d\qq\;d\pp\over (2\p)^4}\
C_\o(\qq+\pp,\qq)\ha_{\pp,\o}\hp^+_{\qq+\pp,\o}\hp^-_{\qq,\o}\;,
\cr\cr
&&\AA_-(\a,\ps)
=
\sum_{\o,\o'=\pm}\int\! {d\qq\;d\pp\over (2\p)^4}\
D_{-\o}(\pp)\n_{\o,\o'}(\pp)
\ha_{\pp,\o}\hp^+_{\qq+\pp,\o'} \hp^-_{\qq,\o'} 
\nn\eea
and,  for
$\pp,\qq\in (\g^{l-1},\g^{N+1})$,
$$C_\o(\qq,\pp) =
[\c_{l,N}^{-1}(\pp)-1] D_\o(\pp) -[\c_{l,N}^{-1}(\qq)-1] D_\o(\qq)\;.
$$
By explicit computation one can check 
\be\lb{13}
{\dpr \HH_{l,N}\over \dpr\ha_{\pp,\o}}(0,J,\h)=R_{\o,l,N}(\pp;J,\h)-D_{-\o}(\pp)\sum_{\o'}\n_{\o,\o'}(\pp)
{\partial \WW_{l,N}\over\partial \hJ_{\pp,\o'}}(J,\h)
\nn\ee
The fundamental issue behind the Adler-Bell-Jackiw anomaly corresponds,
in the viewpoint of our RG scheme, to the following fact:
although  $C_\o(\qq,\pp)$ is zero for $\pp,\qq\in [\g^l,\g^N]$ and 
point-wise vanishing in $(\g^{l-1},\g^{N+1})$ in the limit of removed
cutoffs, its insertion in the graphs of the perturbation theory,
i.e. $R_{\o,l,N}(\pp;J,\h)$, is not vanishing at all. 
Our result is that the remainder, in the limit of removed cutoffs, can
anyways be computed:
\be\lb{h11}
\lim_{-l,N\to\io}{\dpr \HH_{l,N}\over \dpr\ha_{\pp,\o}}(J,\h,0)
={1\over 4\p} D_{-\o}(\pp) \hJ_{-\pp,\o}\;.
\ee
(Recall that \pref{h11} must be understood as generator of
identities for correlations). This formula gives \pref{wi}.
In order to prove it, we need a multiscale integration of
$\HH_{l,N}$. Define the effective potential on scale $h$, $\AA^{(h)}$, 
such that
\be\lb{ep1}\;
e^{\HH_{l,N}(\a,J,\h)}=\int\!dP_{l,h}(\ps)\ 
e^{\VV^{(h)}(\ps, J, \h) +\AA^{(h)}(\a,J,\h,\ps)}\;,
\ee
(so that $\AA^{(h)}(0,J,\h,\ps)=0$) and, correspondingly, the kernels
of the monomials  of $\AA^{(h)}$ that are linear in $\a$
\be\lb{1n2m}
H^{(1;n;2m)(h)}_{\o;\uo';\uo}(\zz;\uw;\ux,\uy)
\defi
\prod_{i=1}^{n}{\dpr \over \dpr J_{\ww_i,\o'_i}}
\prod_{i=1}^{m}{\dpr \over \dpr \ps^+_{\xx_i,\o_i}}
{\dpr \over \dpr \ps^-_{\yy_i,\o_i}}
{\dpr \AA^{(h)}\over \dpr\a_{\zz,\o}}(0,0,0,0)\;.
\ee
For the results in this paper, we only need $n=0,1$.
Because of the definitions of $\AA_0$ and $\AA_-$, we can have quite
an explicit formula for the Fourier transforms of 
$H^{(1;n;2m)(h)}_{\o;\uo';\uo}$. Consider the identity, at $\a=\h=0$, 
\bea\lb{prch}
{\dpr \AA^{(h)}\over \dpr\ha_{\pp,\o}}=
\int{d\qq\over (2\p)^2} C_\o(\qq+\pp,\qq) 
\hp^+_{\qq,\o}\hp^-_{\pp+\qq,\o}\hskip5em
\cr\cr\cr
+\int{d\qq\over (2\p)^2} C_\o(\qq+\pp,\qq) \lft[
\hp^-_{\qq+\pp,\o}{\dpr \VV^{(h)}\over \dpr\hp^-_{\qq,\o}}\hg_\o(\qq)
-\hg_\o(\qq+\pp){\dpr \VV^{(h)}\over \dpr\hp^+_{\pp+\qq,\o}}\hp^+_{\qq,\o}\rgt]
\cr\cr\cr
+
\sum_{i,j=h}^N\int{d\qq\over (2\p)^2} \hU^{(i,j)}_\o(\qq+\pp,\qq)
{\dpr \VV^{(h)}\over \dpr\hp^+_{\pp+\qq,\o}}
{\dpr \VV^{(h)}\over \dpr\hp^-_{\qq,\o}}
\cr\cr\cr
+
\sum_{i,j=h}^N\int{d\qq\over (2\p)^2} \hU^{(i,j)}_\o(\qq+\pp,\qq)
{\dpr^2 \VV^{(h)}\over \dpr\hp^+_{\pp+\qq,\o}\dpr\hp^-_{\qq,\o}}
\cr\cr\cr
-\sum_{\o''}D_{-\o}(\pp)\n_{\o,\o''}(\pp)
{\dpr \VV^{(h)}\over \dpr\hJ_{\pp,\o''}}\;,
\eea
for
$$
\hU^{(i,j)}_\o(\qq+\pp,\qq)
\defi
C_{\o}(\qq+\pp,\qq)
\hg^{(i)}_\o(\qq+\pp) \hg^{(j)}_\o(\qq)\;.
$$
That suggests the following decomposition of the kernels:
\bea\lb{102m}
\hH^{(1;n;2m)(h)}_{\o;\uo';\uo}(\pp;\uk;\uq)=
\hH^{(1;n;2m)(h)}_{0,\o;\uo';\uo}(\pp;\uk;\uq)+
\sum_{\s=\pm}D_{\s\o}(\pp)\hH^{(1;n;2m)(h)}_{\s,\o;\uo';\uo}(\pp;\uk;\uq)
\cr\cr
+\sum_{\s=\pm}D_{\s\o}(\pp)\hK^{(1;n;2m)(h)}_{\s,\o;\uo';\uo}(\pp;\uk;\uq)\;.
\eea
The first line of \pref{prch} corresponds to uncontracted $\AA_{0}$, therefore is not
included in the kernels. In $\hH^{(1;n;2m)(h)}_{0,\o;\uo';\uo}$ there are the terms generated
by the second line of \pref{prch}, i.e. graphs in which only one
between $\ps^-_{\qq,\o}$ and $\ps^+_{\qq+\pp,\o}$ is contracted;
$\hH^{(1;n;2m)(h)}_{\s,\o;\uo';\uo}$ comes from the third line of
\pref{prch}, i.e. when  $\ps^-_{\qq,\o}$ and $\ps^+_{\qq+\pp,\o}$ are
both contracted but to different graphs. Fourth and fifth line of \pref{prch} are kept together
(because we want to exploit a partial cancellation among them) and
generate $\hK^{(1;n;2m)(h)}_{\s,\o;\uo';\uo}$. 
To explain the sum over $\s$, we  define $\hS_{\o',\o}^{(i,j)}$ such
that 
\be\lb{du}
\c_{l,N}(\pp/2)\hU_\o^{(i,j)}(\qq+\pp,\qq)
=\sum_{\s=\pm}D_{\s\o}(\pp)
\hS_{\s\o,\o}^{(i,j)}(\qq+\pp,\qq)\;
\ee
(we freely add the factor $\c_{l,N}(\pp/2)$ because we are only
interested in the case of fixed $\pp\neq0$),
so that, for example, for the Fourier transform of
$\hK^{(1;n;2m)(h)}_{\s,\o;\uo';\uo}$  we find  
\bea\lb{71ter}\;
K^{(1;n;2m)(h)}_{\s,\o;\uo';\uo}(\zz;\ux;\uy)
=\sum_{i,j=h}^N\int\! d\uu d\ww\;
S^{(i,j)}_{\s\o,\o}(\zz;\uu,\ww)W^{(n;2+2m)(h)}_{\uo';\o,\uo}(\ux;\uu,\ww,\uy)
\cr\cr
-\d_{\s,-1}\sum_{\o''}\int\! d\ww\;
 \n_{\o,\o''}(\zz-\ww)W^{(1+n;2m)(h)}_{\o'',\uo';\uo}(\ww;\ux,\uy)
\eea
This kernel for $(n,2m)=(0,2)$ and
$(1,0)$ is depicted in the the first line of Fig.\ref{p5}
and Fig.\ref{p6}, respectively.

\begin{theorem}
For $|\l|$ small enough and fixed $\pp\neq 0$, 
there exists a $C>1$ and a $\th:0<\th<1$ such that, for any $k:M\le k\le N$
and $\s=\pm$,
\bea\lb{hbh}
&&|\hK^{(1;0;2)(k)}_{\s,\o;\o'}(\pp;\kk)|\le C|\l|\g^{-\th(N-k)}
\cr\cr
&&|\hK^{(1;1;0)(k)}_{\s,\o;\o'}(\pp)-\d_{\o,\o'}\d_{\s,-1}{\d(\pp)\over 4\p}|
\le C \g^{-\th(N-k)}
\eea
and, for any other integer $(n,2m)$, 
\bea\lb{hbh2}
&&|\hK^{(1;n;2m)(k)}_{\s,\o;\uo}(\pp;\uq,\uk)|\le
(C|\l|)^n\g^{(1-m-n)k}\g^{-\th(N-k)}
\eea%
\bea\lb{hbh3}
&&|\hH^{(1;n;2m)(k)}_{\s,\o;\uo}(\pp;\uq,\uk)|\le
(C|\l|)^n\g^{(1-m-n)k}\g^{-\th(N-k)}
\eea
\end{theorem}
{\bf\0Proof.} Note that $\hU^{(i,j)}_\o(\qq+\pp,\qq)$ is zero if none
of $i$ and $j$ is $N$ or $l$. Besides, in the appendix of \cite{[FM]} there is the proof of the
following bound: for any $q$ positive integer, 
there exists a constant $C_{q}>1$ such that 
\bea\lb{61}
|S_{\bar\o,\o}^{(i,j)}(\zz;\xx,\yy)|
\le C_{q} {\g^{i}\over 1+[\g^i|\xx-\zz|]^q}
{\g^{j}\over 1+[\g^j|\yy-\zz|]^q}\;.
\eea
Also, we need the result of \cite{[BFM1]}, 
\be\lb{61b}
\lim_{-l,N\to \io }
\sum_{i,j=l+1}^N
\int\!{d\pp\over (2\p)^2}\
\hS_{-\o,\o}^{(i,j)}(\pp,\pp)
={1\over 4\p}\;.
\ee
We bound $|\hK^{(1;n;2m)(k)}|$ and $|\hH^{(1;n;2m)(k)}|$ with the $L_1$
norm of $K^{(1;n;2m)(k)}$ and $H^{(1;n;2m)(k)}$, respectively. 
The estimate \pref{hbh3} is a straightforward consequence of
\pref{hb}, \pref{hb2}  and \pref{61} for $i$ or $j$ equal to $N$; 
indeed, in the graphs expansion
if $\hH^{(1;n;2m)(k)}_{\s,\o;\uo}$, there is no loop 
to worry about 
other than those ones already in the kernels $W^{(n;2m)(k)}$. 
The estimate \pref{hbh2}, for
$(n,2m)\neq(0,2), (1,0)$,  is simple, because it derives from
\pref{hbh} by standard methods.  
For the estimate of the relevant and marginal kernels, 
\pref{hbh}, we have to take advantage of partial cancellations.
 
Using the expansion for $W^{(0;4)(h)}_{\o,\o'}$ in Fig.\ref{p4}, 
we can expand \pref{71ter} for $(n,2m)=(0,2)$ according to Fig.\ref{p5} 
(we have also used, for the class of graphs (d), the decomposition in  Fig.\ref{p1}).
Consider the case $\s=-$.
\insertplot{420}{225}
{\ins{87pt} {217pt}{$\o$}
\ins{87pt} {199pt}{$\zz$}
\ins{125pt}{189pt}{$\uu$}
\ins{125pt}{227pt}{$\ww$}
\ins{161pt}{204pt}{$\o'$}
\ins{150pt}{191pt}{$\xx$}
\ins{161pt}{231pt}{$\o'$}
\ins{150pt}{226pt}{$\yy$}

\ins{195pt}{214pt}{$-\ {\l\over 4\p}$}

\ins{241pt}{218pt}{$\o''$}
\ins{227pt}{199pt}{$\zz$}
\ins{243pt}{199pt}{$\ww$}
\ins{289pt}{204pt}{$\o'$}
\ins{275pt}{191pt}{$\xx$}
\ins{289pt}{230pt}{$\o'$}
\ins{275pt}{226pt}{$\yy$}

\ins{25pt} {155pt}{$=$}
\ins{52pt} {163pt}{$\o$}
\ins{52pt} {141pt}{$\zz$}
\ins{102pt}{141pt}{$\uu$}
\ins{120pt}{141pt}{$\ww$}
\ins{161pt}{150pt}{$\o'$}
\ins{150pt}{135pt}{$\xx$}
\ins{161pt}{177pt}{$\o'$}
\ins{150pt}{172pt}{$\yy$}
\ins{60pt} {180pt}{(a)}

\ins{195pt}{157pt}{$-\ {\l\over 4\p}$}
\ins{241pt}{163pt}{$\o''$}
\ins{226pt}{141pt}{$\zz$}
\ins{245pt}{141pt}{$\ww$}
\ins{289pt}{150pt}{$\o'$}
\ins{275pt}{135pt}{$\xx$}
\ins{289pt}{177pt}{$\o'$}
\ins{275pt}{172pt}{$\yy$}
\ins{220pt}{180pt}{(b)}

\ins{25pt}  {94pt}{$+$}
\ins{62pt} {101pt}{$\o$}
\ins{62pt}  {83pt}{$\zz$}
\ins{92pt} {119pt}{$\uu$}
\ins{138pt}{117pt}{$\uu'$}
\ins{138pt} {73pt}{$\ww$}
\ins{117pt} {85pt}{$\ww'$}
\ins{161pt}{115pt}{$\o'$}
\ins{150pt}{110pt}{$\yy$}
\ins{161pt} {87pt}{$\o'$}
\ins{150pt} {77pt}{$\xx$}
\ins{60pt} {120pt}{(c)}

\ins{195pt} {94pt}{$+$}
\ins{222pt}{101pt}{$\o$}
\ins{222pt}{83pt}{$\zz$}
\ins{249pt}{69pt}{$\uu$}
\ins{277pt}{68pt}{$\uu'$}
\ins{294pt}{75pt}{$\vv$}
\ins{310pt}{85pt}{$\vv'$}
\ins{303pt}{100pt}{$\ww$}
\ins{259pt}{95pt}{$\ww'$}
\ins{355pt}{87pt}{$\o'$}
\ins{340pt}{77pt}{$\xx$}
\ins{355pt}{115pt}{$\o'$}
\ins{340pt}{110pt}{$\yy$}
\ins{220pt}{120pt}{(d)}

\ins{25pt}{24pt}{$+\ \d_{\o,\o'}$}
\ins{72pt}{30pt}{$\o$}
\ins{72pt}{13pt}{$\zz$}
\ins{112pt}{1pt}{$\ww$}
\ins{152pt}{15pt}{$\o'$}
\ins{142pt}{1pt}{$\xx$}
\ins{137pt}{30pt}{$\ww'$}
\ins{147pt}{51pt}{$\o'$}
\ins{118pt}{50pt}{$\uu=\yy$}
\ins{60pt}{50pt}{(e)}

\ins{178pt}{24pt}{$+\ \d_{\o,\o'}$}
\ins{222pt}{30pt}{$\o$}
\ins{222pt}{13pt}{$\zz$}
\ins{262pt}{1pt}{$\ww$}
\ins{302pt}{15pt}{$\o'$}
\ins{292pt}{1pt}{$\xx$}
\ins{287pt}{30pt}{$\ww'$}
\ins{277pt}{50pt}{$\uu$}
\ins{294pt}{53pt}{$\uu'$}
\ins{340pt}{51pt}{$\o'$}
\ins{326pt}{50pt}{$\yy$}
\ins{220pt}{50pt}{(f)}
}
{p5}{\lb{p5}Graphical representation of $K^{(1;0;2)(k)}_{-,\o,\o'}$}{0}
Fixed the integer $q$ and 
calling $b_j(\xx)\defi C_q\g^j/(1+[\g^j|\xx|]^q)$,
we bound the r.h.s. member 
in the same spirit as in Section \ref{s4.1}; 
though this time we also want to find the  exponential small factor 
$\g^{-\th(N-k)}$.

Let's first consider graphs (a) and (b) together:
\bea\lb{75}
\l \sum_{\o''}\int\! d\uu\;
\lft[\sum_{i,j=k}^N
S^{(i,j)}_{-\o,\o}(\zz;\uu,\uu)
-{\d(\zz-\uu)\over 4\p}\rgt]
\int\!d\ww\; v_{\o,\o''}(\uu-\ww)W^{(1;2)(k)}_{\o'';\o'}(\ww;\xx,\yy)
\eea
Using the identity \pref{idb}, 
for graph (a) we have
\bea\lb{125b}
&&\l\sum_{\o''}\sum_{i,j=k}^N\int\! d\uu d\ww \;
S^{(i,j)}_{-\o,\o}(\zz;\uu,\uu)v_{\o,\o''}(\uu-\ww)
W^{(1;2),(k)}_{\o'';\o'}(\ww;\xx,\yy)
\cr\cr
&&=\l\sum_{\o''}\int\! d\ww\;  v_{\o,\o''}(\zz-\ww)
W^{(1;2),(k)}_{\o'';\o'}(\ww;\xx,\yy)\sum_{i,j=k}^N\int\! d\uu \;
S^{(i,j)}_{-\o,\o}(\zz;\uu,\uu)
\cr\cr
&&+\l
\sum_{p=0,1}\sum_{\o''}\sum_{i,j=k}^N\int\! d\uu \;
S^{(i,j)}_{-\o,\o}(\zz;\uu,\uu)(u_p-z_p)
\cdot\cr\cr
&&\hskip3em\cdot\int_0^1\!d\t\;
\int\! d\ww\; (\dpr_pv_{\o,\o''})(\zz-\ww+\t(\uu-\zz))
W^{(1;2),(k)}_{\o'';\o'}(\ww;\xx,\yy)
\eea
The latter term  of 
the r.h.s. member of \pref{125b}
has the wanted estimate: using that one 
between $i$ and $j$ is on scale $N$,
a bound for its norm is
\be
8|\l|\|W^{(1;2),(k)}_{\o'';\o'}\|
\|\dpr v_{\o,\o''}\|_{L_1}
\|b_N\|_{L_1(w)} \sum_{i=k}^N \|b_i\|_{L_\io}  \le |\l|C_5\g^{-(k-M)}\g^{-(N-k)}\;.
\ee
The former term of 
the r.h.s. member of \pref{125b} - as opposed to what happened for (b3) of
Fig\ref{p3} - 
is not zero, but is compensated by (b): 
\bea\lb{78bis}
\sum_{i,j=k}^N\int\! d\uu \;
S^{(i,j)}_{-\o,\o}(\zz;\uu,\uu)-{1\over 4\p}=
2\sum_{j\le k-1}\int\! d\uu \;
S^{(N,j)}_{-\o,\o}(\zz;\uu,\uu)
\eea
and hence the bound for such a difference is $C_5\g^{-(N-k)}$. The
global bound for (a) and (b) together is therefore
$|\l|C_6\g^{-(N-k)}$.

Graph (c) corresponds to 
\bea
\l\sum_{i,j=k}^N\sum_{\o''}\int\! d\uu d \uu' d\ww d\ww'\;
S^{(i,j)}_{-\o,\o}(\zz;\uu,\ww)
g_\o(\uu-\uu')v_{\o,\o''}(\uu-\ww')
\cdot\cr\cr\cdot
W^{(1;4)(k)}_{\o'';\o,\o'}(\ww';\uu',\ww,\xx,\yy)\;.
\eea
Since either $i$ or $j$ has to be $N$, and 
because of the bound \pref{61},
the norm of (c) is bounded by 
\bea
|\l|\sum_{m=k}^{N}\sum_{i,j=k}^{*N}\int\!  d \xx  d \uu' d\ww d\ww'\;
|W^{(1;4)(k)}_{\o'';\o,\o'}(\ww';\uu',\ww,\xx,\yy)|
\cr\cdot
\int\!d\zz d\uu \;
b_i(\zz-\ww) b_j(\zz-\uu)|v_{\o,\o''}(\uu-\ww') g^{(m)}_\o(\uu-\uu')|
\eea
where $*N$ reminds that at least one between $i$ and $j$ has to be $N$.
As in the previous section, we bound the second line as follows:
\bea
\|b_N\|_{L_1}\  \|b_j\|_{L_{3/2}}\ \|v_{\o,\o'}\|_{L_3} \ \|g^{(m)}_\o\|_{L_\io}
\quad{\rm for}\quad i=N,\ m\le j
\cr\cr
\|b_N\|_{L_1}\ \|b_j\|_{L_\io}\  \|v_{\o,\o'}\|_{L_3}\ \|g^{(m)}_\o\|_{L_{3/2}}
\quad{\rm for}\quad i=N,\ j< m
\cr\cr
\|b_i\|_{L_1}\  \|b_N\|_{L_{3/2}}\ \|v_{\o,\o'}\|_{L_3} \ \|g^{(m)}_\o\|_{L_\io}
\quad{\rm for}\quad j=N,\ m\le i
\cr\cr
\|b_i\|_{L_\io} \  \|b_N\|_{L_1}\ \|v_{\o,\o'}\|_{L_3} \ \|g^{(m)}_\o\|_{L_{3/2}}
\quad{\rm for}\quad j=N,\ i< m
\eea
and hence we get the bound,
for $0\le \th \le1/3$, $C_3>1$ 
\be\lb{222} 
\int\!d\zz d\uu \;
b_i(\zz-\ww) b_j(\zz-\uu)|v_{\o,\o''}(\uu-\ww')
g^{(m)}_\o(\uu-\uu')|\le C_3
 \g^{-{4\over 3}(k-M)}\g^{-\th (N-k)}\g^k
\ee
(with $C_3\to\io$ if $\th \to 1/3$). Using \pref{hb2} for
$W^{(1;4)(k)}_{\o'';\o,\o'}$  we have for graph (c) the bound  
$|\l|C_6 \g^{-{4\over 3}(k-M)}\g^{-\th (N-k)}$. Graph (d) is bounded in the same way as (c) (in fact (d) was
distinguished from (c) only for enumeration reasons, whereas
topologically they are the same). We find
\bea
\sum_\s\int\!  d \xx  d \uu' d\ww d\ww'\;
|W^{(1;2)(k)}_{\o'';\o}(\uu;\ww',\uu')g_\o(\vv-\ww)v_{\o,\s}(\ww-\vv')
W^{(1;2)(k)}_{\s;\o'}(\vv';\xx,\yy)|
\cr\cdot
|\l|^2\sum_{m=k}^{N}\sum_{i,j=k}^{N*}\int\!d\zz d\uu \;
b_i(\zz-\ww) b_j(\zz-\uu)|v_{\o,\o''}(\uu-\ww') g^{(m)}_\o(\uu-\uu')|\;;
\nn\eea
then, with the aid of \pref{hb} and \pref{222}, we find a bound of the
type $|\l|C_3 \g^{-{4\over 3}(k-M)}\g^{-\th (N-k)}$.  For (e) and (f), by
a simple argument, we have the bound 
$$
4|\l|\|W^{(1;2)(k)}_{\o'';\o'}\|
\lft[1+\|g_\o\|_{L_1}\|W^{(0;2)(k)}_{\o}\|\rgt]\|v_{\o,\o'}\|_{L_3}
\sum_{i,j=k}^{*N}
\|b_i\|_{L_1}\|b_j\|_{L_{3/2}}\;,
$$
that less than $|\l|C_3 \g^{-{4\over 3}(k-M)}\g^{-\th (N-k)}$.
Now consider the case $\s=+$. The graph expansion of
$K^{(1;0;2)(k)}_{+,\o;\o'}$ is given again by Fig.\ref{p5};  the
only  differences is that  the graph (b) is missing (that because of
the $\d_{\s,-1}$ in \pref{71ter}). 
Hence a bound can be obtained 
with the same above argument, with only one important difference: 
the contribution that in the previous analysis were compensated 
by (b) now are zero by symmetries. Indeed,  
calling $\kk^*$ the rotation of $\kk$ of $\p/2$ and 
since $\hS^{(i,j)}_{\bar\o,\o}(\kk^*,\pp^*)
=-\o\bar\o\hS^{(i,j)}_{\bar\o,\o}(\kk,\pp)$,
in place of \pref{78bis}, in this case we have:
\be\lb{79}
\sum_{i,j=k}^N\int\! d\uu \;
S^{(i,j)}_{\o,\o}(\zz;\uu,\uu)=
\sum_{i,j=k}^N\int\! {d\kk\over (2\p)^2} \;
\hS^{(i,j)}_{\o,\o}(\kk,-\kk)=0
\ee
The proof of the first of \pref{hbh} is completed. 
We now consider the second.
%
Expand $W^{(1;2)(h)}_{\o';\o}$ as in Fig.\ref{p2}, and obtain the
decomposition of Fig.\ref{p6}. In particular, class (e) comes from
the kernel $w^{(1;2)}_{\o,\o'}$ that is is darker bubble of
Fig.\ref{p2}; while for (d) and (f) we also used the identity in
Fig.\ref{p1} to extract a further wiggly line. It is also worth
stressing that (e) is not included in (a), because by construction  
$W^{(2;0)(h)}_{\o,\o'}(\ww,\xx)$ does not contain $\d_{\o,\o'} \d(\xx-\ww)$. 
\insertplot{420}{225}
{\ins{87pt} {217pt}{$\o$}
\ins{87pt} {199pt}{$\zz$}
\ins{125pt}{189pt}{$\uu$}
\ins{125pt}{227pt}{$\ww$}
\ins{161pt}{220pt}{$\o'$}
\ins{156pt}{201pt}{$\xx$}

\ins{195pt}{212pt}{$-\ {1\over 4\p}$}

\ins{241pt}{220pt}{$\o''$}
\ins{227pt}{199pt}{$\zz$}
\ins{243pt}{199pt}{$\ww$}
\ins{289pt}{220pt}{$\o'$}
\ins{281pt}{199pt}{$\xx$}

\ins{25pt} {155pt}{$=$}
\ins{52pt} {163pt}{$\o$}
\ins{52pt} {141pt}{$\zz$}
\ins{102pt}{141pt}{$\uu$}
\ins{120pt}{141pt}{$\ww$}
\ins{161pt}{166pt}{$\o'$}
\ins{156pt}{141pt}{$\xx$}
\ins{60pt} {180pt}{(a)}

\ins{193pt}{156pt}{$-\ {1\over 4\p}$}
\ins{241pt}{166pt}{$\o''$}
\ins{216pt}{141pt}{$\zz=\uu$}
\ins{245pt}{141pt}{$\ww$}
\ins{289pt}{166pt}{$\o'$}
\ins{281pt}{141pt}{$\xx$}
\ins{220pt}{180pt}{(b)}

\ins{25pt}  {94pt}{$+$}
\ins{62pt} {101pt}{$\o$}
\ins{62pt}  {83pt}{$\zz$}
\ins{92pt} {119pt}{$\uu$}
\ins{138pt}{117pt}{$\uu'$}
\ins{138pt} {73pt}{$\ww$}
\ins{117pt} {85pt}{$\ww'$}
\ins{161pt}{104pt}{$\o'$}
\ins{156pt}{83pt}{$\xx$}
\ins{60pt} {120pt}{(c)}

\ins{193pt} {94pt}{$+$}
\ins{222pt}{101pt}{$\o$}
\ins{222pt}{83pt}{$\zz$}
\ins{249pt}{69pt}{$\uu$}
\ins{277pt}{68pt}{$\uu'$}
\ins{294pt}{75pt}{$\vv$}
\ins{310pt}{85pt}{$\vv'$}
\ins{303pt}{100pt}{$\ww$}
\ins{259pt}{95pt}{$\ww'$}
\ins{350pt}{104pt}{$\o'$}
\ins{345pt}{83pt}{$\xx$}
\ins{220pt}{120pt}{(d)}

\ins{25pt}{24pt}{$+\ \d_{\o,\o'}$}
\ins{72pt}{30pt}{$\o$}
\ins{72pt}{13pt}{$\zz$}
\ins{132pt}{34pt}{$\o'$}
\ins{130pt}{13pt}{$\xx$}
\ins{60pt}{50pt}{(e)}

\ins{178pt}{24pt}{$+\ \d_{\o,\o'}$}
\ins{222pt}{30pt}{$\o$}
\ins{227pt}{13pt}{$\zz$}
\ins{255pt}{-1pt}{$\uu$}
\ins{280pt}{-3pt}{$\uu'$}
\ins{312pt}{33pt}{$\o'$}
\ins{300pt}{3pt}{$\ww$}
\ins{308pt}{16pt}{$\xx$}
\ins{270pt}{26pt}{$\ww'$}
\ins{220pt}{50pt}{(f)}
}
{p6}{\lb{p6}Decomposition of the class $K^{(1;1;0)(h)}_{-,\o;\o'}$}{0}
It is evident that graphs of classes (a), (b), (c) and (d) can be bounded as the
graphs in homonym classes in Fig.\pref{p5}: the only difference is one external
wiggly line in place of two external fermion lines, but that does not
change the power counting, nor the topology of the graph.  A  bound
for graph (f) is:
\bea
|\l|\int\!d\uu'd\ww'd\ww\;
|W^{(1;2)(k)}_{\o'',\o}(\ww';\uu',\xx) g_\o(\ww-\xx)|
\cr\cr
\sum_{i,j=k}^{*N}
\int\!d\zz d\uu\;
 b_i(\zz-\uu)b_j(\zz-\xx) |v_{\o,\o''}(\uu-\ww') g_\o(\uu-\uu')|\;.
\eea
By \pref{222} we have the bound $|\l|C_3 \g^{-\th(N-k)}$.

The only graph that is not bounded by the exponential small factor is
(e). In fact, this kernel is finite; and to cancel it we need the 
to subtract $\d_{\s,-1}/(4\p)$, see Fig.\ref{p7}.
\insertplot{420}{25}
{\ins{100pt}{4pt}{$\zz$}
\ins{150pt}{4pt}{$\xx$}

\ins{192pt}{16pt}{$-\ {1\over 4\p}$}
\ins{235pt}{4pt}{$\zz=\xx$}
}
{p7}{\lb{p7}Cancellation of graph (e) in Fig.\ref{p6} for $\s=-1$ 
with the factor
$1/(4\p)$.}{0}
The difference equals \pref{78bis} for $\s=-1$, and \pref{79} for
$\s=+1$. Therefore also the second of \pref{hb2} is proved. \qed

We still have to consider a last kind of kernels, 
$\hH^{(1;n;2m)(k)}_{0,\o;\uo}$. They can easily bounded, supposing
$\pp\neq0$ and finite. Anyways to extract the small factor one might
need the IR integration also: indeed, either the small factor comes
from a contraction  of $\AA_0$ on
scale $N$, or on scale $l$; the (simple) details are in \cite{[BFM1]}.

As consequence of the analysis in this section, by the
argument in \cite{[BFM1]},  in the limit of removed cutoffs, 
the correlations generated by $\HH$ satisfy the identities 
generated by \pref{h11}. 
\subsection{Closed Equation}\lb{s4.3}

From  the Wick theorem, see \pref{wt1}, we find the \sde\ equation
\be\lb{sde3}
{\dpr^2\WW_{l,N}\over \dpr \hh^+_{\kk,\o}\dpr \hh^-_{\kk,\o}}(0,0)
=\hg^{[l,N]}_{\o}(\kk)\lft[1+\l\sum_{\o'}\int\!{d\pp\over (2\p)^2}
\hv_{\o,\o'}(\pp) 
{\dpr^3\WW_{l,N}\over \
\dpr \hJ_{\pp,\o'}\dpr \hh^+_{\kk+\pp,\o}\dpr \hh^-_{\kk,\o}}(0,0)\rgt]
\ee
that, in the limit of removed cutoffs, would be equivalent to
\pref{SDE}; we do not take the limit now, though.  Using \pref{12} and
\pref{13},  we find
\bea\lb{88}
\sum_{\o'',\o'}\Big[D_\o(\pp)(\hv^{-1}(\pp))_{\o,\o''}-D_{-\o}(\pp)
{\d_{\o,\o''}\over 4\p}\Big]\hv_{\o'',\o'}(\pp)
{\dpr\WW_{l,N}\over \dpr \hJ_{\pp,\o'}}(0,\h)
\cr\cr
\hskip3em
=B_{\pp,\o}(0,\h)
+
{\dpr\HH_{l,N}\over \dpr \ha_{\pp,\o}}(0,0,\h)
\eea
where the inverse of the matrix $\hv_{\o,\o'(\pp)}$ is
$$
(\hv^{-1}(\pp))_{\o,\o'}=(\pp^2+\m^2)\d_{\o,-\o}
-{1\over2 }\lft(1-\a+{\s\over \pp^2}\rgt)\lft[(\pp^0)^2-\o\o'(\pp^1)^2-i(\o+\o')\pp^0\pp^1\rgt]\;.
$$
As done to prove \pref{vec2} and \pref{ax2}, use that 
$$
\sum_\o D_\o(\pp)(\hv^{-1}(\pp))_{\o,\o'}=
\lft(\a\pp^2+\m^2-\s\rgt)D_{-\o'}(\pp)
$$
$$
\sum_\o\o D_\o(\pp)(\hv^{-1}(\pp))_{\o,\o'}=
-\o'\lft(\pp^2+\m^2\rgt)D_{-\o'}(\pp)
$$
to obtain a more explicit form of  \pref{88}
\bea\lb{83}
\sum_{\o'}\hv_{\o,\o'}(\pp)
{\dpr\WW_{l,N}\over \dpr \hJ_{\pp,\o'}}(0,\h)
=\sum_{\o'} M_{\o,\o'}(\pp)
\lft[B_{\pp,\o'}(0,\h)
+
{\dpr\HH_{l,N}\over \dpr \ha_{\pp,\o'}}(0,\h)\rgt]
\eea
for $M_{\o,\o'}(\pp)$  the Fourier transform of
$-\dpr_{\o} F(\xx)+\o'\dpr_{\o} F_5(\xx)$. Plug \pref{83} into 
\pref{sde3} and obtain an equation that, in the limit of removed cutoff,
equals \pref{SDE2}, but for a remainder term, 
\be\lb{remT}
\sum_{\o'}\int{d\pp\over (2\p)^2}
M_{\o,\o'}(\pp){\dpr^2\HH_{l,N}\over \dpr \ha_{\pp,\o'}\dpr \hh^+_{\kk+\pp,\o}}(0,\h)\;.
\ee
We have to prove that, in the limit of removed cutoffs, \pref{remT} is
vanishing. To this purpose, as in the previous section, we introduce a
new functional integral
\bea
e^{\TT_{\e,l,N}(\b,\h)} = 
\int\! dP_{l,N}(\ps) e^{\VV(\ps,0,\h)
+\BB_{\e,0}(\ps,\b)-\BB_{\e,-}(\ps,\b)}
\eea
for
\bea\lb{80}
\BB_{\e,0}(\psi,\b)= \sum_{\o} \int\! {d\kk\;d\pp\;d\qq \over
  (2\p)^6}\;
M_{\o,\e\o}(\pp) C_{\e\o}(\qq+\pp,\qq)\;
\hb_{\kk,\o} \hp^-_{\kk+\pp,\o} \hp^+_{\qq+\pp,\e\o}
\hp^-_{\qq,\e\o}\;,
\eea
\be\lb{80a}
\BB_{\e,-}(\psi,\b)= \sum_{\o,\o'} \int\!{d\kk\;d\pp\;d\qq\over (2\p)^6}\;
M_{\o,\o'}(\pp) D_{-\o'}(\pp)\n_{\o',\e\o}(\pp)\;\hb_{\kk,\o} \hp^-_{\kk+\pp,\o} \hp^+_{\qq+\pp,\e\o}
\hp^-_{\qq,\e\o}\;,
\ee
Therefore we find:
$$
{\dpr \TT_{\e,l,N}\over\dpr \b_{\kk,\o}}(0,\h)
=
\int{d\pp\over (2\p)^2}
M_{\o,\e\o}(\pp){\dpr^2\HH_{l,N}\over \dpr \ha_{\pp,\e\o}\dpr
  \hh^+_{\kk+\pp,\o}}(0,\h)\;.
$$
We now perform a multiscale integration of $\TT_{\e,l,N}(\b,0)$.
Define $\BB^{(h)}(\b,\h,\ps)$, the effective potential on scale $h$,
to be such that  
\be\lb{ep3}\;
e^{\TT_{\e,l,N}(\b,\h)}=\int\!dP_{l,h}(\ps)\ 
e^{\VV^{(h)}(\ps,0,\h) +\BB_\e^{(h)}(\b,\h,\ps)}\;,
\ee
and correspondingly, the kernels of the monomials of $\TT_\e^{(h)}$ that
are linear in $\b$:
\be\lb{kert}
T^{(2m;2)(h)}_{\e;\uo,\o}(\ux,\uy;\uu,\vv)
\defi
\prod_{i=1}^{m}{\dpr \over \dpr \ps^+_{\xx_i,\o_i}}
{\dpr \over \dpr \ps^-_{\yy_i,\o_i}}
{\dpr^2 \BB_\e^{(h)}\over \dpr\b_{\uu,\o} \dpr\ps^-_{\vv,\o}}(0,0,0)\;.
\ee
To make more explicit the kernels $T^{(2m;2)(h)}_{\e;\uo,\o}$ we use
the following identity at $\b=\h=0$:
\bea
{\dpr \BB^{(h)}_\e\over\dpr \b_{\kk,\o}}
=\sum_{i,j=h}^N\int{d\pp d\qq\over (2\p)^4}
M_{\o,\e\o}(\pp)\hU^{(i,j)}_{\e\o}(\qq+\pp,\qq) 
{\dpr^2\VV^{(h)}\over
\dpr \hp^+_{\qq,\e\o}\dpr \hp^-_{\qq+\pp,\e\o}}
{\dpr \VV^{(h)}\over \dpr \hh^+_{\kk+\pp,\o}}
\cr\cr\cr
\lb{prct}+\sum_{\o'}\int{d\pp \over (2\p)^2}
M_{\o,\o'}(\pp)D_{-\o'}(\pp)\n_{\o',\e\o}(\pp)
{\dpr \VV^{(h)}\over \dpr \hJ_{\pp,\e\o}} 
{\dpr \VV^{(h)}\over \dpr \hh^+_{\kk+\pp,\o}}
\cr\cr\cr
+\sum_{i,j=h}^N\int{d\pp d\qq\over (2\p)^4}
M_{\o,\e\o}(\pp)\hU^{(i,j)}_{\e\o}(\qq+\pp,\qq)\hg_\o(\kk+\pp) 
{\dpr^3 \VV^{(h)}\over \dpr \hp^+_{\kk+\pp,\o}
\dpr \hp^+_{\qq,\e\o}\dpr \hp^-_{\qq+\pp,\e\o}}
\cr\cr\cr
+\sum_{\o'}\int{d\pp \over (2\p)^2}
M_{\o,\o'}(\pp)D_{-\o'}(\pp)\n_{\o',\e\o}(\pp)\hg_\o(\kk+\pp)  
{\dpr^2 \VV^{(h)}\over \dpr \hp^+_{\kk+\pp,\o}
\dpr \hJ_{\pp,\e\o}}
\cr\cr\cr
-\int{d\pp d\qq\over (2\p)^4}
M_{\o,\e\o}(\pp)C_{\e\o}(\qq+\pp,\qq) 
 \ e^{-\VV^{(h)}}{\dpr\over \dpr \hh^+_{\kk+\pp,\o}}
\lft[e^{\VV^{(h)}}{\dpr \VV^{(h)}\over \dpr\hh^+_{\qq,\e\o}}
     {\dpr \VV^{(h)}\over \dpr\hh^-_{\qq+\pp,\e\o}}\rgt]
\eea
We decompose the kernels as follows:
$$
\hT^{(2m;2)(h)}_{\e;\uo,\o}(\uq;\kk)=\hT^{(2m;2)(h)}_{0,\e;\uo,\o}(\uq;\kk)
+\hT^{(2m;2)(h)}_{1,\e;\uo,\o}(\uq;\kk)
+\hT^{(2m;2)(h)}_{2,\e;\uo,\o}(\uq;\kk)\;;
$$
in $T^{(2m;2)(h)}_{0,\e;\uo,\o}(\ux;\uu,\vv)$ we collect the term
generated by the last line of \pref{prct}; in
$T^{(2m;2)(h)}_{1,\e;\uo,\o}(\ux;\uu,\vv)$ we included the terms
generated by the third and fourth line of \pref{prct}; finally,
$T^{(2m;2)(h)}_{2,\e;\uo,\o}(\ux;\uu,\vv)$ is related  to the first
two lines of \pref{prct}.
\begin{theorem}
For $|\l|$ small enough, 
there exists a $C>1$ and a $\th:0<\th<1$ such that, for any $k:M\le
k\le N$, $r=1,2$  and $\e=\pm$
\be\lb{hbt2}
|\hT^{(2m;2)(k)}_{r,\e;\uo,\o}(\uq,\kk)|\le C^m |\l|\g^{(1-m)k} e^{-\th(N-k)}
\ee
\end{theorem}

{\bf\0Proof.} 
Again we use the $L_1$ norm of $T^{(2m;2)(k)}$ as upper bound of $|T^{(2m;2)(k)}|$.
In the cases $r=1,2$ the proof of \pref{hbt2} is a direct consequence
of \pref{hbh}, \pref{hbh2}. 
Indeed, for $r=2$, there is not new loop in the graph
expansion of  $\hT^{(2m;2)(k)}_{r,\e;\uo,\o}$ w.r.t. the graph
expansion of $ \hK^{(1;0;2m)(k)}_{\s,\o;\uo}$ for $\s=\pm$. Whereas
for $r=1$ there is only one loop more, that can be easily bounded:
for 
$u_{\o,\e\o}^{(\s)}(\xx)$ the Fourier transform of
$M_{\o,\e\o}(\pp)D_{\s\e\o}(\pp)$
$$
T^{(2m;2)(k)}_{1,\e;\o}(\uw;\xx,\yy)
=\sum_\s\int\!d\zz d\uu \;
u^{(\s)}_{\o,\e\o}(\xx-\zz) g_\o(\xx-\uu) 
K^{(1;0;2m+2)(k)}_{\s;\e\o,\o'} (\zz;\uw,\uu,\yy)
$$
\insertplot{430}{55} 
{\ins{95pt}{31pt}{$\o$}
\ins{103pt}{23pt}{$\xx$}
\ins{202pt}{31pt}{$\o$}
\ins{193pt}{23pt}{$\yy$}
\ins{116pt}{51pt}{$\zz$}
\ins{150pt}{3pt}{$\uu$}

\ins{225pt}{31pt}{$-$}

\ins{254pt}{31pt}{$\o$}
\ins{283pt}{23pt}{$\xx$}
\ins{295pt}{48pt}{$\zz$}
\ins{344pt}{31pt}{$\o$}
\ins{333pt}{23pt}{$\yy$}
\ins{333pt}{23pt}{$\yy$}
\ins{295pt}{3pt}{$\uu$}
} {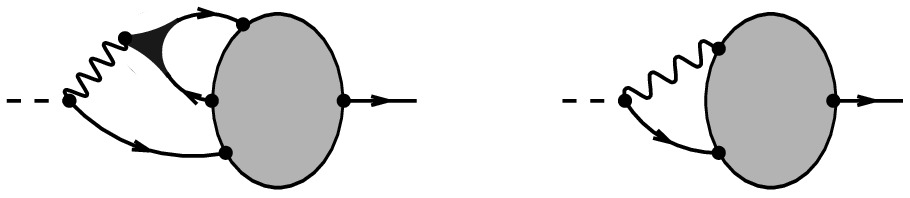}{\lb{p9b}Graphical representation
of $T^{(0;2)(k)}_{1,\e;\o}$. The dashed line represents the external
fermion field $\b$.}{0}
Note that the bound for
$\|u_{\o,\e\o}^{(\s)}\|_{L_p}$ is essentially the same of
$\|v_{\o,\o'}\|_{L_p}$; therefore, using \pref{hbh}, we obtain the bound
\be
2\|u^{(\s)}_{\o,\e\o}\|_{L_3} \sum_{j=k}^N \|g_\o^{(j)}\|_{L_{3/2}}
\|K^{(1;0;2m+2)(k)}_{\s;\e\o,\o'}\| \le C |\l| \g^k \g^{-{4\over3}(k-M)}
\g^{-\th(N-k)}\;.
\ee
That completes the proof of the theorem. \qed

We shall now analyze the last kind of kernel left,
$T^{(2m;2)(k)}_{0,\e;\o}(\uq;\kk)$. We further expand the last line of 
\pref{prct}
\bea\lb{prctt}
\int{d\pp d\qq\over (2\p)^4}
M_{\o,\e\o}(\pp)C_{\e\o}(\qq+\pp,\qq) 
{\dpr\VV^{(h)}\over \dpr \hh^+_{\kk+\pp,\o}}
{\dpr \VV^{(h)}\over \dpr\hh^+_{\qq,\e\o}}
{\dpr \VV^{(h)}\over \dpr\hh^-_{\qq+\pp,\e\o}}
\cr\cr\cr
-\int{d\pp d\qq\over (2\p)^4}
M_{\o,\e\o}(\pp)C_{\e\o}(\qq+\pp,\qq) \hg_{\e\o}(\qq)
{\dpr^2 \VV^{(h)}\over \dpr \hh^+_{\kk+\pp,\o}\dpr\hp^+_{\qq,\e\o}}
\hp^+_{\qq+\pp,\e\o}
\cr\cr\cr
+\int{d\pp d\qq\over (2\p)^4}
M_{\o,\e\o}(\pp)C_{\e\o}(\qq+\pp,\qq) \hg_{\e\o}(\qq+\pp)
{\dpr\VV^{(h)}\over \dpr \hh^+_{\kk+\pp,\o}} 
\hp^-_{\qq,\e\o}
{\dpr \VV^{(h)}\over \dpr\hp^-_{\qq+\pp,\e\o}}
\cr\cr\cr
+\int{d\pp d\qq\over (2\p)^4}
M_{\o,\e\o}(\pp)\hU_{\e\o}(\qq+\pp,\qq)\hg_\o(\kk+\pp) 
{\dpr\over \dpr \hp^+_{\kk+\pp,\o}}
\lft[{\dpr \VV^{(h)}\over \dpr\hp^+_{\qq,\e\o}}
     {\dpr \VV^{(h)}\over \dpr\hp^-_{\qq+\pp,\e\o}}\rgt]
\eea
In the terms generated by the first line each of 
$\hp^-_{\kk+\pp}$, $\hp^+_{\qq+\pp,\e\o}$ and $\hp^-_{\qq,\e\o}$
is contracted with a different kernel $\hW^{(0,2m)(k)}$ (if any);  
therefore this term  is bounded with \pref{hb}, \pref{hb2}; 
the small factor can be extracted only if at least one between
$\hp^+_{\qq+\pp,\e\o}$ and  $\hp^-_{\qq,\e\o}$ is contracted;
otherwise it comes from the IR integration, as in \cite{[BFM1]}.

In the terms generated by the second and third line, one between
$\hp^+_{\qq+\pp,\e\o}$ and $\hp^-_{\qq,\e\o}$ is  not contracted;
anyways in the graphical representation there is a loop that is not
included in the kernels $\hW^{(0;2m)(k)}$, with momentum $\pp$.
\insertplot{430}{50} 
{
\ins{82pt}{28pt}{$\o$}
\ins{82pt}{18pt}{$\kk$}
\ins{97pt}{37pt}{$\pp$}
\ins{152pt}{54pt}{$\e\o$}
\ins{155pt}{42pt}{$\kk_1$}
\ins{105pt}{1pt}{$\kk+\pp$}
\ins{173pt}{16pt}{$\o'$}
\ins{173pt}{35pt}{$\o'$}

\ins{214pt}{25pt}{$+$}

\ins{242pt}{28pt}{$\o$}
\ins{242pt}{18pt}{$\kk$}
\ins{257pt}{37pt}{$\pp$}
\ins{312pt}{54pt}{$\e\o$}
\ins{315pt}{42pt}{$\kk_1$}
\ins{265pt}{1pt}{$\kk+\pp$}
\ins{333pt}{16pt}{$\o'$}
\ins{333pt}{35pt}{$\o'$}
} 
{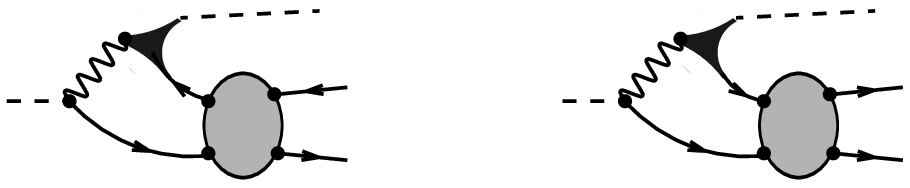}{\lb{p13}Graphical representation
of possible terms generated by 
the second line of \pref{prctt}}{0}
Its
explicit expression is
\bea
\int{d\pp\over (2\p)^2}\;
M_{\o,\e\o}(\pp)
\Big[\big(1-f_N(\kk_1-\pp)\big)-
\big(\c_{l,N}^{-1}(\kk_1)-1\big)\hg_\o(\kk_1-\pp)\Big]
\cr\cr\cdot
\hg_\o(\kk+\pp)\hW^{(0;2m+2)(k)}_{\o,\e\o\uo}(\kk+\pp,\kk_1-\pp,\uk)
\eea
The addend proportional to $\big(1-f_N(\kk_1-\pp)\big)$ has the
constraint that $|\pp|\ge c\g^N$; then  $|M(\pp)|\le c\g^{-3N}$, and
the bound is $C\g^{-N}\|g_\o\|_{L_1}\|W^{(0;2m+2)(k)}\|\le 
C_M\g^{-(N-k)}\g^{-2(k-M)}\g^{-(1-m)k}$;  whereas the addend
proportional to $\big(\c_{l,N}^{-1}(\kk_1)-1\big)$ will be contracted
on scale $l$ (otherwise is zero), so obtaining the exponentially small
factor, see \cite{[BFM1]}. 

Finally, from the fourth line we obtain terms in which
$\hp^-_{\kk+\pp,\o}$ is contracted; and also 
both $\hp^+_{\qq+\pp,\e\o}$ and $\hp^-_{\qq,\e\o}$ are contracted, but
one of them is linked to the same kernel as  $\hp^-_{\kk+\pp,\o}$.  
We find
\bea
\sum_\s\sum_{i,j=k}^N
\int\!d\zz d\ww d\ww' d\uu d\uu'\;
u^{\s}_{\o,\e\o}(\zz-\ww)
S^{(i,j)}_{\s\o,\o}(\ww-\uu,\ww-\uu')
g_\o(\zz-\ww')
\cr\cr
W^{(0;2m_1)(k)}(\uu',\ux)
W^{(0;2m_2+2)(k)}(\ww',\uu,\ux')
\eea
that is bounded by
\bea
4\|u^{\s}_{\o,\e\o}\|_{L_3} \|g_\o\|_{L_{3/2}} 
\|b_N\|_{L_1} \sum_{j=k}^N\|b_j\|_{L_1}\|W^{(0;2m_1)(k)}\|
\|W^{(0;2m_2+2)(k)}\|
\cr\cr
\le C\g^{(2-m_1-m_2)} \g^{-(N-k)}\g^{-{4\over 3}(k-M)} \;.
\eea
The consequence of the analysis in this section is that, using the
argument in \cite{[BFM1]},  the correlations generated by \pref{remT}
are vanishing when cutoffs are removed. 
\vskip2em
{\bf\0Acknowledgments.} I thank V.Mastropietro for suggesting the
problem.
{\it This material is based upon work supported by the National
  Science Foundation under agreement No. DMS-0635607. Any opinions,
  findings and conclusions or recommendations expressed in this
  material are those of the author and do not necessarily reflect the
  views of the National Science Foundation.}
\pagina

\appendix
\section{Proof of the graphical identities}\lb{a1}
We remind that the derivative in the Grassmann variables $\ps^+$,
$\h^+$ and $\z^+$ are taken from the left, while the derivatives in
$\ps^-$, $\h^-$ and $\z^-$ are taken from the right.   
The definition of $\VV^{(k)}$ is given by \pref{ep0}. Accordingly, the
relation between $\VV^{(k)}$ and $\VV$ is
\be\lb{ep2}\;
\VV^{(k)}(\ps,J,0)=\ln\int\!dP_{k+1,N}(\z)\ 
e^{\VV(\ps+\z, J, 0)}\;.
\ee
and we have the two identities:
\bea
\lb{aprc1}&&{\dpr e^{\VV^{(k)}}\over \dpr \ps^+_{\xx,\o}}(\ps,J,0)
=J_{\xx,\o}{\dpr e^{\VV^{(k)}}\over\dpr \h^+_{\xx,\o}}(\ps,J,0)
\cr\cr
&&\phantom{********}+\l\sum_{\o'}\int\!d\ww\ 
v_{\o,\o'}(\xx-\ww)
{\dpr^2e^{\VV^{(k)}}\over\dpr J_{\ww,\o'} \dpr \h^+_{\xx,\o}}(\ps,J,0)\;.
\\
\cr\cr
\lb{aprc2}&&{\dpr \VV^{(k)}\over \dpr J_{\xx,\o}}(\ps,J,\h)
=-e^{-\VV^{(k)}(\ps,J,\h)} 
{\dpr^2e^{\VV^{(k)}}\over
  \dpr\h^+_{\xx,\o}\dpr\h^-_{\xx,\o}}(\ps,J,\h)
\eea
 Moreover the {\it Wick theorem} for 
$dP(\z)$
Gaussian mean values gives
\bea\lb{wt1}
\int\!dP_{k+1,N}(\z)\ \z_{\xx,\o}^{-\e} F(\z)=\e\int\!d\uu\ 
g^{[k+1,N]}_\o(\xx-\uu)
\int\!dP_{k+1,N}(\z)\  {\dpr F(\z)\over \dpr \z_{\uu,\o}^\e}
\eea
this identity is straightforward for
$F=\exp\{\sum_\xx
\z^+_{\xx,\o}\h^-_{\xx,\o}+\sum_\xx \h^+_{\xx,\o}\z^-_{\xx,\o}\}$ 
and so is also for any formal power series with
even numbers of fields. Therefore
\bea\lb{wt2}
&&{\dpr e^{\VV^{(k)}}\over\dpr \h^\e_{\xx,\o}}(\ps,J,\h)
=\int\!dP_{k+1,N}(\z)\ 
\lft(\ps_{\xx,\o}^{-\e}+\z_{\xx,\o}^{-\e}\rgt) e^{\VV(\ps+\z,J,\h)}
\cr\cr 
&&=
\ps_{\xx,\o}^{-\e}e^{\VV^{(k)}} 
+\e\int\! d\uu\ 
g^{[k+1,N]}_\o(\xx-\uu) 
{\dpr e^{\VV^{(k)}}\over \dpr \ps_{\uu,\o}^\e}(\ps,J,\h)\;.
\eea
We plug the identity for $\e=+$ into \pref{aprc1} and get \pref{prc1}.
Also, since $g_\o^{[k+1,N]}(0)=0$, from \pref{wt2} and \pref{aprc2} we obtain:
\bea
{\dpr e^{\VV^{(k)}}\over \dpr J_{\xx,\o}}(\ps,J,\h)
=-
\ps^-_{\xx,\o}
{\dpr e^{\VV^{(k)}}\over \dpr\h^-_{\xx,\o}}(\ps,J,\h)
\hskip5em
\cr\cr
-\int\! d\uu\ 
g^{[k+1,N]}_\o(\xx-\uu) 
{\dpr ^2
e^{\VV^{(k)}}\over \dpr \ps_{\uu,\o}^+\dpr\h^-_{\xx,\o}}(\ps,J,\h)
\eea
that gives \pref{prc2}.

\end{document}